# Viable Supply Chain Network Design: Machine Learning-Derived Chance-Constrained Programming

Mohammad Rohaninejad[a], Behdin Vahedi-Nouri[b], Elham Jelodari Mamaghani[c], Mehdi Foumani[d], Olga Battaia[e,1*]

[a] *Czech Institute of Informatics, Robotics, and Cybernetics, Czech Technical University in Prague, Prague, Czech Republic*

[b] *School of Industrial Engineering, College of Engineering, University of Tehran, Tehran, Iran*

[c] *Institute of Sustainable Business and Organisations Sciences and Humanities Confluence Research Center-UCLY, ESDES, Lyon, France*

[d] *Department of Industrial Engineering, Hainan Bielefeld University of Applied Sciences, Hainan, China*

[e] *KEDGE Business School, Bordeaux, France*

## Abstract:

This paper investigates a viable two-echelon supply chain network design problem with unreliable facilities subject to disruptions. Unlike existing studies that consider supply chain echelons in isolation, the proposed models explicitly capture cross-echelon disruptions and quantify the value of incorporating such interdependencies. Network viability is obtained by combining resilience through backup reassignment, agility through mobile facilities, and environmental considerations such as emission limits. Together, these elements help maintain demand satisfaction across both echelons and support long-term network performance. Two mixed-integer programming formulations are developed. The first formulation is a scenario-based formulation and the second is implicit formulation which both minimize expected fixed and service costs. To handle probabilistic service requirements, the implicit formulation integrates a machine learning–enhanced chance-constrained programming approach. In this framework, intractable capacity chance constraints are approximated by learned linear cuts that apply a 95% service confidence level. These cuts are trained using several classification methods, including logistic regression, L1-regularized logistic regression, stochastic gradient descent, the perceptron algorithm, and logistic regression with a regularization parameter of 0.1. The best-performing classifier is then selected as a surrogate model. To improve scalability, two fix-and-relax heuristics are developed for the implicit formulation, while a

[1*] Corresponding author.

*E-mail address:* olga.battaia@kedgebs.com(O. Battaia).

sample average approximation (SAA) method is used for the scenario-based formulation. Computational experiments demonstrate that the implicit formulation proposes a computationally efficient and high-quality alternative to the scenario-based formulation. Furthermore, the proposed heuristics and SAA approach effectively address medium- and large-scale instances, delivering high-quality solutions within acceptable computational times.

## 1. Introduction and literature review

Supply chain network design problems (SCNDPs) involve determining the optimal locations of a number of facilities, with finite or infinite capacity, among demand points, as well as determining the flow of materials between them. SCNDPs have long been central to supply chain management because they determine how infrastructure supports the flow of goods and services (Celik & Genevois, 2020; Kalczynski et al., 2025) by achieving a balance between first-stage establishment costs, second-stage service and transportation costs. Classical SCNDP formulations usually assume that facilities remain available, clients are always served, and demand is completely satisfied. Although these assumptions make the model easier to handle, they do not capture the reality of supply chains, where disruptions occur frequently (Rohaninejad et al., 2018a; Cheng et al., 2021a; Maliki et al., 2025).

Every year, numerous firms experience unexpected disruptions within their supply chain networks. These kinds of disruptions can be caused by many different factors such as natural disaster, power failure, transfer of ownership, operational failure, and strikes by workers (Cheng et al., 2021b; Cheng et al., 2024). Therefore, there has been an increasing trend in scientific literature to study the different types of uncertainties affecting supply chain performance. Initially, the focus was more on receiver-side uncertainty, where the amount of demand is stochastic (Nazemi et al., 2022). They also examined distribution-side uncertainty, particularly variations in transportation costs and delivery lead times (Huang et al., 2012). In addition to *receiver-side* and *distribution-side* uncertainties, *provider-side* uncertainty, such as random facility failures or capacity losses, is a critical yet relatively underexplored factor in supply chain resilience (Rohaninejad et al., 2018b). Because SCNDPs are long-term structural decisions, they cannot be easily adjusted in response to disruptions, and repeatedly reconfiguring the network is neither feasible nor theoretically justified (Wang and Ouyang, 2013). Thus, integrating facility disruption impacts on the design phase becomes crucial for making supply chains more resilient.

However, being only resilient does not provide a complete solution to cope with the challenges posed by complex supply chains. Resilient network design models, which strive to achieve service delivery even in the face of disruptions, neglect two key factors. First is adaptability and ability to respond rapidly to changing conditions. The second one is sustainability and incorporating environmental effects. As a result, a supply chain may remain operational under disruptions yet lack responsiveness or fail to meet sustainability expectations. To address these limitations, the concept of supply chain viability has been proposed as a more comprehensive perspective. Rather than focusing only on resilience, viability integrates resilience, agility, and sustainability (Ivanov, 2022; Ivanov, 2023; Ivanov & Tu, 2025; Padovano & Ivanov, 2025; Dolgui et al., 2025). In this context, resilience guarantees continuity through redundancy and reassignment mechanisms (Cui et al., 2010; Hosseini & Ivanov, 2022; Katsaliaki et al., 2022). Agility reflects how quickly a system can be reconfigured and respond to disruptions through adaptive mechanisms such as replacement. Finally, sustainability considers environmental impacts in supply chain decision-making (Govindan, 2025).

More importantly, viability emphasizes the ability of supply chains to remain functional not only under anticipated disruptions but also in the presence of unknown and unforeseen disturbances (Ivanov, 2022; Ivanov, 2023; Ivanov & Tu, 2025; Padovano & Ivanov, 2025; Ivanov, 2025). In line with this perspective, recent studies have begun to operationalize viability in dynamic disruption contexts. For example, Broekaert et al. (2025) evaluated viability using longitudinal service levels together with resilience metrics such as time to recovery. Similarly, Rinaldi and Grosse (2026) showed, through a simulation-based approach, that dynamic reconfiguration strategies can enhance supply chain viability under disruptions.

However, there are still some limitations. The first limitation is in the fact that although agile management has attracted increasing interest (Akram et al., 2024), it has yet to be formally integrated within the field of SCNDP. Another problem is that the environmental impacts have been studied separately from disruption risk and have yet to be considered together in disruption-aware SCNDP optimization models. This means that existing research efforts fail to provide a unified optimization model which simultaneously addresses resilience, agility, and sustainability in disruption-aware supply chain network design.

In order to position this study within the emerging viability literature, it is important to differentiate it from current conceptual and empirical contributions. Earlier studies that focus on resilience, like Hosseini and Ivanov (2022), created probabilistic frameworks using Bayesian Networks to examine how disruptions spread through supply networks and impact system performance. These studies mainly concentrate on disruption dynamics and performance assessment, not on decisions related to network design.

Recent research has developed the concept of supply chain viability within a comprehensive analysis. For example, Padovano and Ivanov (2025) proposed a multidimensional viability framework using structural equation modeling. They empirically validated the suggested model by focusing on firm-level capabilities. Although these studies suggested significant conceptual and empirical insights, they did not directly address optimization-based network design under disruptive conditions. The present study focuses on the viability within a formal optimization framework by introducing the viable two-echelon supply chain network design problem (VTE-SCNDP). In this context, resilience is represented by the reassignment of backup facilities, agility by the deployment of mobile facilities, and the environmental aspect by emission constraints. This study advances the viability literature in multi-echelon supply chain design by integrating the three dimensions into a mixed-integer programming model. It also develops customized solutions to effectively address the proposed problem.

A review of the SCNDP literature shows that most classical formulations considered multi-echelon structures without incorporating the viability dimension (Pan and Naji, 2013; Petridis, 2015). The studies on viability and reliability have focused on a single-echelon supply chain, without addressing the propagation of disruptions from one layer to other layers of the supply chain. Although Thomas and Mahanty (2021) and Putthakosa and Trung Luong (2025) studied two-echelon structures and disruption in one echelon, they ignored the potential for disruption to propagate to the other echelons across interconnected layers of the network. This narrow focus is problematic because supply chains are inherently interconnected, and disruptions in one echelon create inter-echelon demand dependencies that propagate through the network, adversely affecting downstream operations and overall performance. Because the operational requirements of the second echelon are endogenously driven by the first, treating echelons independently risks suboptimal or even infeasible network designs. Addressing this limitation requires modeling frameworks that explicitly integrate resilience, agility, and sustainability to ensure the long-term survival of the integrated supply chain.

Although studies addressing simultaneously resilience and viability under uncertainty in a multi-echelon structure have increased recently, they are predominantly application-specific. Furthermore, they have not provided a comprehensive optimization framework for multi-echelon network design under facility disruptions. For instance, Yılmaz et al. (2025a) proposed a two-echelon stochastic programming model within machine learning for humanitarian supply chain design. Yılmaz et al. (2025b) developed a unified robust stochastic programming approach to improve supply chain viability through integrated warehouse design decisions. This issue is important because the function of supply chains depends on the integration of all echelons. If a problem occurs in one echelon, it can quickly propagate throughout the supply chain and disrupt the system. To have a strong network design, it's vital to include the effects of disruption across the whole system, not just each part in isolation. We need mathematical models that see the whole picture and combine strength, flexibility, and sustainability to handle disruptions.

However, the computational complexity associated with ensuring reliable and viable SCNDPs poses substantial challenges. Even the deterministic versions of these problems, including their capacitated variants, are NP-hard (Irawan and Jones, 2019; Tang & Wu, 2024; González et al., 2025). As a result, a wide range of heuristics and metaheuristics has been developed to improve scalability (Caserta and Quiñonez Rico, 2009; Fernandes et al., 2014; Irawan and Jones, 2019; González et al., 2021; Turkeš et al., 2021). Although these approaches have been widely applied, their performance often varies across problem settings, and many remain highly problem-specific (Chalupa and Nielsen, 2019).

Table 1 compares existing studies with this work and highlights an important research gap. Even though there's been progress in addressing uncertainty in supply chain network design problems (SCNDPs), most studies haven't fully incorporated the concept of viability, especially when facility disruptions occur. To address this gap, this study introduces a two-echelon supply chain network design problem (VTE-SCNDP) that clearly includes the main parts of viability in one optimization model. In particular, resilience is modeled through the reassignment of backup facilities. Agility is captured through the deployment of mobile facilities, defined as temporary, rapidly deployable units that provide service when regular facilities become unavailable (Alarcon-Gerbier and Buscher, 2022; Chen and Yu, 2016; Liu et al., 2019; Lei et al., 2016). In practice, such facilities correspond to deployable capacity units such as mobile medical facilities and field hospitals in emergency logistics (Kelen et al., 2015; Yücel et al., 2020), trailer-mounted power generation units used for rapid service restoration in energy systems (Panteli and Mancarella, 2015), and temporary or modular processing and maintenance units in industrial systems (Smith, 2010). These facilities are temporary capacity providers during disruptions to fixed facilities. The model also makes sure environmental impact is managed by setting clear emission limits. The integrated structure facilitates the evaluation of whether coordinated decision-making across multiple echelons enhances overall system performance and long-term viability in comparison to independent, echelon-based designs.

This study introduces two novel mixed-integer programming (MIP) formulations. The first formulation is an implicit formulation (IF) and the second is scenario-based formulation (SBF). The main contribution is the integration of machine learning within the optimization framework to improve both realism and computational tractability. The IF applies a machine learning–based chance-constrained programming approach to replace intractable capacity chance constraints with learned linear cuts. To address large-scale computational complexity, two tailored fix-and-relax heuristics are developed for the IF, and a sample-average approximation (SAA) algorithm is used for the corresponding SBF. To the best of the authors' knowledge, this study is the first to employ fix-and-relax heuristics and machine learning–based constraint learning to viable multi-echelon supply chain network design problems (SCNDPs). Computational results show that the proposed approaches ensure network viability while maintaining tractability for complex supply chain structures.

The remaining sections of the paper are organized as follows. Section 2 defines the problem and its associated assumptions. Sections 3 and 4 present the SBF and IF formulations, respectively. Section 5 explains how machine-learning-derived linear cuts replace the capacity chance constraints. Section 6 describes the proposed fix-and-relax heuristics. Section 7 provides computational results. Finally, Section 8 of this paper outlines conclusions and future research directions.

**Table 1.** Summary of the related literature

| Reference | Agility | Resiliency | Sustainability | Uncertainty | Multi echelon | Solution approach |
|---|---|---|---|---|---|---|
| Broekaert et al. (2025) | - | X | - | - | - | simulation based game theory |
| Caserta and Quiñonez Rico (2009) | - | - | - | - | - | metaheuristic |
| Cui et al. (2010) | - | X | - | X | - | lagrangian relaxation |
| Huang et al. (2012) | - | - | - | - | - | DC |
| Li et al. (2013) | - | X | - | X | - | lagrangian relaxation |
| Pan and Naji (2013) | X | X | - | X | X | lagrangian relaxation |
| Wang and Ouyang (2013) | - | X | - | X | - | game theory |
| Fernandes et al. (2014) | - | - | - | - | X | metaheuristic |
| Petridis (2015) | - | - | - | X | X | scenario-base |
| Bardossy and Raghavan (2017) | - | - | - | X | - | heuristic |
| Fischetti et al. (2017) | - | - | - | - | - | benders decomposition |
| Rohaninejad et al. (2018b) | - | X | - | X | X | heuristic |
| Chalupa and Nielsen (2019) | - | - | - | - | - | metaheuristic |
| Irawan and Jones (2019) | - | - | - | - | X | metaheuristic |
| Cheng et al. (2021a) | - | X | - | X | - | C&CG algorithm |
| Cheng et al. (2021b) | - | X | - | X | - | C&RG |
| González et al. (2021) | - | - | - | - | X | metaheuristic |
| Thomas and Mahanty (2021) | - | - | - | X | X | dynamic simulation |
| Turkeš et al. (2021) | - | X | - | X | X | metaheuristic |
| Ivanov (2020) | X | X | X | X | X | conceptual modeling |
| Nazemi et al. (2022) | - | X | - | X | - | heuristic |
| Ivanov (2023) | X | X | - | X | X | conceptual modeling |
| Akram et al. (2024) | X | - | - | - | X | SEM |
| Cheng et al. (2024) | - | X | - | X | X | scenario-wise DRO |
| Tang and Wu (2024) | - | X | - | X | X | LR+heuristic |
| González et al. (2025) | - | - | X | - | X | metaheuristic |
| Kalczynski et al. (2025) | - | - | - | - | - | SNOPT |
| Maliki et al. (2025) | - | X | - | X | X | metaheuristic |
| Padovano and Ivanov (2025) | X | X | X | X | X | SEM |
| Putthakosa and Trung Luong (2025) | - | X | - | X | X | CTMC |
| Yılmaz et al. (2025a) | - | X | - | X | X | SP + GA + ML |
| Yılmaz et al. (2025b) | - | X | - | X | X | hybrid metaheuristics+ML |
| Rinaldi and Grosse (2026) | X | X | - | X | X | simulation |
| **Our study** | **X** | **X** | **X** | **X** | **X** | **fix and relax/SAA heuristics** |

## 2. Problem definition

In this study, we address the design of a viable two-echelon supply chain network design problem (VTE-SCNDP) that remains functional, adaptable, and environmentally responsible in the face of uncertainty. Viability is achieved by incorporating three essential capabilities into the network design: resilience, agility, and environmental impacts. The facilities at both levels are exposed to the risk of disruption with two operational levels: a) active; and b) inactive. In active mode, the facilities are fully available, while inactive facilities cannot provide any services to clients and are therefore out of reach. The aim of solving this problem is to design a viable network for a two-echelon supply chain so as to minimize the expected total fixed costs and service costs (including the expected transportation costs and costs incurred for failing to meet client demand).

Each facility in the first echelon operates simultaneously as a service provider to end-users and as a client to the facilities in the second echelon. Demand in the first echelon is known and fixed, while demand in the second echelon is derived endogenously, based on upstream assignments. Specifically, the demand of each second-echelon node is a function of the total demand allocated to it from the first echelon. Therefore, it is necessary to examine two-echelon systems from an integrated perspective to enhance the viability of the whole system. It is assumed that, in the event of a disruption, the flow of demand from one echelon to the next may be interrupted. In such cases, if a node is unable to receive the necessary supply from within the network (i.e., from its upstream echelon), it must source the required demand externally. This external sourcing incurs a higher cost, modeled as a penalty cost.

In the case of disruptions that interrupt the normal flow of goods, resilience is achieved by ensuring that clients can be reassigned to alternative facilities when their current assigned service points become unavailable. This ability to redirect demand helps maintain continuity of operations and prevents service breakdowns during failure events. By anticipating potential disruptions and embedding flexibility into the network design, the model ensures that demand can still be met through alternative paths, supporting resilience in the supply chain under uncertainty.

To maintain service responsiveness and agility, an acceptable response time is imposed for assignments to the regular facilities. In situations where no backup facility is available within the allowable response time, the model permits the deployment of temporary mobile facilities. These mobile units can be quickly established in strategic locations to serve affected clients during emergencies when regular facilities are unavailable or fall short. By doing so, the network retains its ability to respond flexibly under time-sensitive conditions. Figure 1 illustrates a schematic representation of the problem, featuring supply and demand nodes in both echelons, as well as the deployment of mobile facilities for enhanced agility.

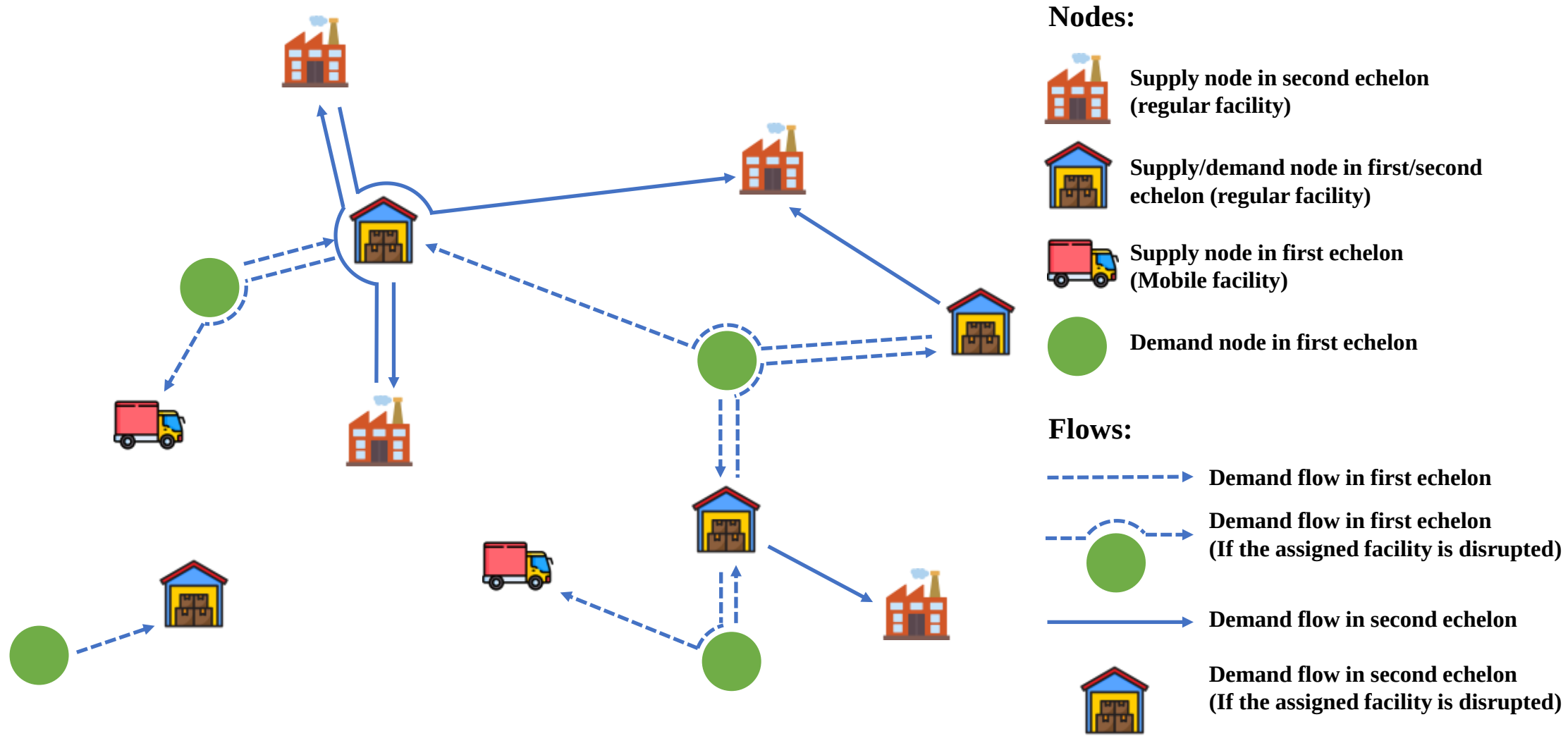

**Figure 1.** Graphical representation of the VTE-SCNDP

Additionally, the model takes into account the environmental impact of transportation activities throughout the supply chain. Each client-to-facility assignment contributes to a cumulative emissions total, calculated based on travel distances and vehicle emission factors. To ensure that the design remains sustainable in terms of environmental impact, a hard constraint is imposed on the total expected emissions. This constraint limits the ecological footprint of the supply chain while supporting long-term operational goals. Other assumptions of the problem are as follows:

- The potential sites for establishing regular facilities are predefined and discrete.
- Facility failures at both echelons are assumed to occur independently.
- The problem is a single-product model.
- The partial or fractional allocation of demand to a facility is not allowed.
- There is a fixed cost to establish both the regular and mobile facility.
- There is a transportation cost to allocate a demand node to a facility node.
- There is a penalty cost for supplying demand from outside the network (unmet demand within the network).
- The maximum number of facilities that can be established in each location is one.
- The network is two-echelon, and all facilities in each echelon are the same (e.g., all facilities are wholesalers or retailers)
- The problem parameters, such as demand in the first echelon, distances, failure probabilities, fixed costs, and transportation costs, are specified.

- Mobile facilities are considered only for serving end users, i.e., clients in the first echelon of the supply chain.
- All demand must be assigned to a regular facility under normal conditions. Mobile facilities are only deployed in response to disruptions when the assigned regular facility is unavailable.
- Mobile facilities operate as temporary, self-contained units during disruptions and do not rely on upstream material flow through the network.
- Each echelon must include a predetermined number of regular facilities, limited by a maximum allowable number defined in the network design.
- Regular facilities are assumed to have sufficient capacity to fully meet the demand assigned to them.
- Each mobile facility can serve at most a predefined number of demand nodes simultaneously.
- A facility cannot serve a client if the travel time between them exceeds a predefined threshold.
- Mobile facilities are assumed to be fully reliable and are not subject to disruptions.
- Each client-facility assignment contributes to overall emissions based on transport distance and emission factors.

We assume that regular facilities have sufficient capacity to satisfy all assigned demand when they are not disrupted. Similar problems with unlimited capacity assumption are widely studied by researchers (De Armas et al., 2017; Yu & Zhang, 2018; Ramshani et al., 2019). This assumption is adopted to control our model complexity. Then, we can maintain a clear focus on the viability dimension, which constitutes the core objective of this study. From a practical standpoint, capacity sufficiency can be achieved through multiple mechanisms. For instance, facilities may operate under integrated production–distribution configurations that allow them to adjust output in response to demand variations (see Martins et al., 2021). Moreover, capacity can be scaled through the allocation of additional resources when demand increases. Such assumptions are commonly adopted in settings involving non-perishable and bulk products, where capacity expansion is operationally feasible and does not require stringent environmental constraints.

At last, it's important to consider that the suggested model, necessarily, is not limited to predefined disruption scenarios. Although disruption risks are included in the mathematical model, incorporating multi-level backup assignments and mobile facilities provides a generalized response mechanism that does not depend on the exact realization of disruptions. In this sense, the model can be adapted to the network design. This enables the system to respond effectively even to unforeseen and previously unobserved disruptions. Hence, the proposed framework supports supply chain viability by allowing the network to maintain its functionality under both known and unknown sources of uncertainty.

## 3. Scenario-based formulation (SBF)

The indices, parameters, and variables of the SBF are as follows:

***Model indices***

| | |
|---|---|
| $i, i'$ | Client index |
| $j, j'$ | Potential site index for the regular facilities |
| $k$ | Potential site index for the mobile facilities in the first echelon |
| $l$ | Echelon index; $(l = 1,2)$ |
| $s$ | Scenario index |

***Model parameters***

| | |
|---|---|
| $d_i$ | Demand quantity of client $i$ in the first echelon |
| $\gamma_l$ | The maximum number of regular facilities must be established in echelon $l$ |
| $s_{l,j}$ | The fixed cost of opening a regular facility in site $j$ of echelon $l$ |
| $\delta_k$ | The fixed cost of opening a mobile facility at site $k$ |
| $c_{l,i,j}$ | The transportation cost per unit demand of client $i$ in echelon $l$ that is serviced by a regular facility $j$; this cost is a function of distance and increases with greater travel distance. |
| $c'_{i,k}$ | The transportation cost per unit demand of client $i$ in the first echelon that is serviced by a mobile facility at site $k$ |
| $f_{l,i,j}$ | The distance between client $i$ in echelon $l$ and regular facility in site $j$ |
| $f'_{ik}$ | The distance between client $i$ in the first echelon and the mobile facility in site $k$ |
| $t_{i,j}$ | The transportation time of client $i$ in the first echelon to regular facility $j$ |
| $t'_{ik}$ | The transportation time of client $i$ in the first echelon that is serviced by mobile facility $k$ |
| $h_{l,i}$ | The penalty cost per unit of client $i$ demand in echelon $l$ if its demand is not met |
| $\theta_j$ | Conversion coefficient of demand volume from the first echelon to the second echelon for facility $j$ in the first echelon $(1 \le \theta_j \le 2)$ |
| $e_l$ | Emission rate per unit demand per unit distance (e.g., kg $CO_2$ per unit × km) in echelon $l$ |
| $E_{max}$ | The maximum allowable expected transportation-related emissions across all disruption scenarios |
| $T_{max}$ | The maximum allowable travel time for the client in the first echelon |
| $q_s$ | Probability of scenario $s$ |
| $b_{l,j,s}$ | A binary parameter such that: |

$b_{l,j,s} = 1$ ; If facility $j$ in echelon $l$ is active and functional under scenario $s$
$b_{l,j,s} = 0$ ; Otherwise

$MC_k$ Maximum number of clients that can be assigned to mobile facility $k$

$I_l$ Total number of clients in echelon $l$

$J_l$ Total number of potential sites for regular facilities in echelon $l$

$K$ Total number of potential sites for mobile facilities in the first echelon

$G$ A positive large number

***Model variables***

$y_{l,j}$ A binary variable such that:
$y_{l,j} = 1$ ; if regular facility $j$ is open in echelon $l$
$y_{l,j} = 0$ ; Otherwise

$\mu_{k,s}$ A binary variable such that:
$\mu_{k,s} = 1$ ; if mobile facility $k$ is open under scenario $s$
$\mu_{k,s} = 0$ ; Otherwise

$x_{i,j}$ A binary variable such that:
$x_{i,j} = 1$ ; If facility $j$ is the primary regular facility assigned to client $i$
$x_{i,j} = 0$ ; Otherwise

$X_{l,i,j,s}$ A binary variable such that:
$X_{l,i,j,s} = 1$ ; If client $i$ assigned to regular facility $j$ at echelon $l$ under scenario $s$
$X_{l,i,j,s} = 0$ ; Otherwise

$M_{i,k,s}$ A binary variable such that:
$M_{i,k,s} = 1$ ; If client $i$ at the first echelon assigned to mobile facility $k$ under scenario $s$
$M_{i,k,s} = 0$ ; Otherwise

$Z_{l,i,s}$ A binary variable such that:
$Z_{l,i,s} = 1$ ; If the demand of client $i$ at echelon $l$ is not met under scenario $s$
$Z_{l,i,s} = 0$ ; Otherwise

$u_{i,j,s}$ A positive variable; specifies the demand of client $i$ provided by facility $j$ at the second echelon under scenario $s$

$v_{i,s}$ A positive variable; specifies the demand of client $i$ at the second echelon that is not met under scenario $s$

The mathematical model of the VTE-SCNDP based on the SBF is as follows:

$$min \sum_{l} \sum_{j \le J_l} s_{l,j} y_{l,j} + \sum_{s} \sum_{k \le K} \delta_k \, \mu_{k,s} \, q_s + \sum_{s} \left( \sum_{i \le I_1} d_i \left( \sum_{j \le J_1} c_{1,i,j} \, X_{1,i,j,s} + \sum_{k \le K} c'_{i,k} \, M_{i,k,s} \right) + \sum_{i \le I_2} \sum_{j \le J_2} c_{2,i,j} \, u_{i,j,s} \right) q_s \\ \sum_{s} \left( \sum_{i \le I_1} h_{1,i} d_i \, Z_{1,i,s} + \sum_{i \le I_2} h_{2,i} v_{i,s} \right) q_s \quad (1)$$

*s.t.*

$$1 \le \sum_{j \le J_l} y_{l,j} \le \gamma_l \qquad \forall \, l \quad (2)$$

$$\sum_{j \le J_1 | t_{i,j \le T_{max}}} x_{i,j} = 1 \qquad \forall\, i \le I_1 \tag{3}$$

$$x_{i,j} \le y_{1,j} \qquad \forall\, i \le I_1;\ j \le J_1 \tag{4}$$

$$X_{1,i,j,s} \ge\ x_{i,j}\, b_{1,j,s} \qquad \forall\, i \le I_1;\ j \le J_1; s \tag{5}$$

$$\sum_{j \le J_1} X_{1,i,j,s} + \sum_{k \le K} M_{i,k,s} + Z_{1,i,s} = 1 \qquad \forall\, i \le I_1;\ \ s \tag{6}$$

$$\sum_{j \le J_2} X_{2,i,j,s} + Z_{2,i,s} \le y_{1,i} \qquad \forall\, i \le I_2;\ \ s \tag{7}$$

$$\sum_{j \le J_2} u_{i,j,s} + v_{i,s} = \sum_{i' \le I_1} \left( d_{i'}\, \theta_{j'} X_{1,i',j',s} \right) \qquad \forall\, i \le I_2;\ \ j' = i\, ; s \tag{8}$$

$$u_{i,j,s} \le G\, X_{2,i,j,s} \qquad \forall\, i \le I_2;\ j \le J_2;\ s \tag{9}$$

$$v_{i,s} \le G\, Z_{2,i,s} \qquad \forall\, i \le I_2;\ s \tag{10}$$

$$\sum_{i \le I_l} X_{l,i,j,s} \le G\, b_{l,j,s}\, y_{l,j} \qquad \forall\, l\, ;\ j \le J_l;\ s \tag{11}$$

$$\sum_{i \le I_1} M_{i,k,s} \le MC_k\, \mu_{k,s} \qquad k \le K;\ s \tag{12}$$

$$\sum_{j \le J_1} t_{i,j}\, X_{1,i,j,s}\ + \sum_{k \le K} t'_{kj}\, M_{i,k,s} \le T_{max} \qquad \forall\, i \le I_1;\ \ s \tag{13}$$

$$\sum_{s} \left( \sum_{i \le I_1} d_i\, e_1 \left( \sum_{j \le J_1} f_{1,i,j}\, X_{1,i,j,s} + \sum_{k \le K} f'_{kj}\, M_{i,k,s} \right) + \sum_{i \le I_2} \sum_{j \le J_2} f_{2,i,j}\, u_{i,j,s} e_2 \right) . q_s\ \le E_{max} \tag{14}$$

Equation (1) represents the objective function, which minimizes the total expected cost across all disruption scenarios. This includes fixed costs of opening regular and mobile facilities, transportation costs for serving client demands, and penalties for any unmet demand. Constraint (2) guarantees the minimum and maximum number of regular facilities that must be established in each echelon. Constraints (3)–(5) ensure that each client is assigned to one main regular facility in the first echelon. Constraint (5) ensures that the assignment is maintained in all scenarios where the facility is operational, except in cases where the client's demand is supposed to be unmet. Constraints (6) and (7) ensure that all customer demand in the first and second echelons, respectively, is either satisfied by the supply chain or penalized if unmet. Because demand is not necessarily assigned to all second-echelon facilities, Constraint (7) is formulated as an inequality. Constraint (8) is then introduced to correctly capture the demand flow from the second echelon to the first echelon. Specifically, Constraint (8) links the flow originating from second-echelon facilities to the aggregate demand assigned to them by first-echelon customers through a conversion coefficient. Constraints (9) and (10) eliminate partial assignment by enforcing that each client in the second echelon is assigned to exactly one facility (or declared unmet). Constraint (11) ensures that no client can be assigned to a facility that is not active and open under the given scenario. Constraint (12) limits the number of clients that can be assigned to any mobile facility, reflecting the capacity restrictions of temporary mobile facilities. This ensures mobile units are not overloaded during deployment. Constraint (13) introduces the travel-time-based agility restriction, prohibiting assignments to mobile facilities if the travel time between the client and the mobile facility exceeds a predefined threshold. This constraint guarantees responsiveness limits in the event of disruptions. Constraint (14) introduces a hard-environmental constraint on the total expected emissions from transportation. The emissions are computed based on client demand, travel distances, and an emission rate, and must remain below a specified threshold $E_{max}$ across all scenarios.

## 4. Implicit formulation

While the SBF provides a detailed and accurate representation of the disruption environment by explicitly modeling all possible failure scenarios, its scalability becomes increasingly challenging as the number of clients, facilities, and scenarios grows. The exponential increase in problem size can significantly impact computational performance, especially in large-scale instances. To address this issue, we develop an alternative implicit formulation (IF) that avoids direct enumeration of scenarios by modeling backup relationships between clients and facilities. In this approach, each client is pre-assigned to a sequence of backup facilities based on their priority and proximity. When a facility fails, the client is served by the next available facility in its predefined backup list. This design captures the essence of resilience while maintaining computational tractability, making the IF more suitable for efficiently solving larger instances of the problem. Figure 2–a illustrates how resiliency is incorporated into the network design through the concept of backup facility assignments. However, for the sake of simplicity, we consider a simplified model where we assume that mobile facilities can only serve as second-level (i.e., $r = 2$) backup options. Figure 2–b displays a schematic representation of this simplification. As a result, the solution space of the implicit formulation may differ slightly from that of the SBF, potentially representing a restricted subset of the broader solution space encompassed by the SBF.

$$\sum_{s' \neq s} \left( \sum_{j' \neq j} X_{1,i,j',s'} + \sum_{k' \neq k} M_{i,k',s'} \right) \leq (2 - M_{i,k,s} - x_{i,j})G \qquad \forall i \leq I_1; j \leq J_1; k \leq K; s \; | b_{1,j,s} = 0 \tag{15}$$

Specifically, this means that Constraint (15) is imposed within the SBF. This constraint ensures that if client $i$ is assigned to a mobile facility $k$ in a scenario $s$, where its primary regular facility $j$ is inactive (i.e., $b_{1,j,s} = 0$), then no alternative assignment is allowed in any other scenario $s' \neq s$, where $b_{1,j,s'} = 0$.

At the same time, this simplification facilitates the formulation of more efficient IF constraints that address the capacity and fixed establishment cost of mobile facilities. To assess the validity and practical value of this approach, we compare the optimal solutions obtained from the IF with those derived from the SBF, thereby evaluating whether the simplification introduced by the IF is justifiable.

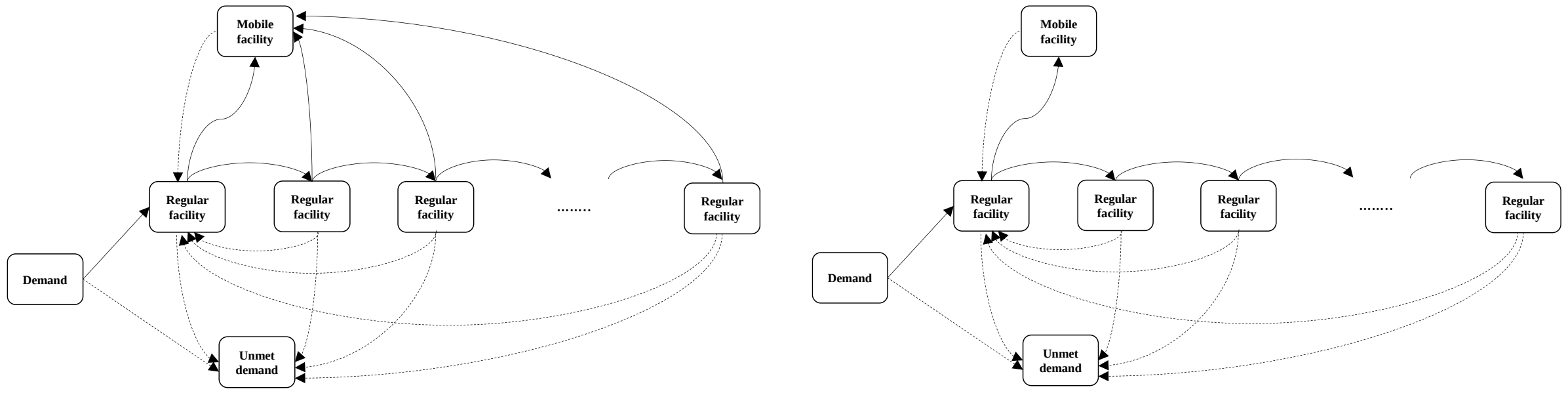

a) Original model b) Simplified model

**Figure 2.** Graphical representation of backup-based establishing resiliency in the original and simplified models

The IF model has been shown to solve single-echelon uncapacitated facility location problems with facility disruption more efficiently than scenario-based approaches (Shen et al., 2011). However, this type of formulation has limitations when it is extended from single-echelon to multi-echelon problems. The main challenge lies in the difficulty of tracking demand flows across multiple echelons, which significantly increases model complexity. In this paper, for the first time, we develop an implicit formulation for a VTE-SCNDP that successfully captures demand flows in both the first and second echelons. The indices, parameters, and variables of the proposed IF are as follows:

***Model indices***

$r$ Multiple backup levels for the allocation of clients to facilities

***Model parameters***

| | |
|---|---|
| $p_l$ | Failure probability of facilities in echelon $l$ |
| $R_l$ | Maximum allowed number of backup facilities for clients in echelon $l$ |
| $\alpha_{n,k}$ | Coefficient from the trained linear classifier corresponding to the $n^{th}$ feature of mobile facility $k$ |
| $\beta_{n,k}$ | Intercept from the trained linear classifier corresponding to the $n^{th}$ feature of mobile facility $k$ |

***Model variables***

| | |
|---|---|
| $x_{i,j,r}$ | A binary variable such that:<br>$x_{i,j,r} = 1$ ; if facility $j$ is the $r^{th}$ level backup facility of client $i$ in the first echelon<br>$x_{i,j,r} = 0$ ; Otherwise |
| $m_{i,k}$ | A binary variable such that:<br>$m_{i,k} = 1$ ; if mobile facility $k$ is the $2^{th}$ level backup facility of client $i$ in the first echelon<br>$m_{i,j} = 0$ ; Otherwise |
| $z_{i,r}$ | A binary variable such that:<br>$z_{i,r} = 1$ ; if client $i$ in the first echelon has $(r-1)$ backup facilities, but has no $r^{th}$ backup facility<br>$z_{i,r} = 0$ ; Otherwise |
| $\rho_k$ | Probability of establishing a mobile facility in site $k$ |
| $Q_{i',i,j,r,r'}$ | A binary variable such that:<br>$Q_{i',i,j,r,r'} = 1$ ; if the demand of client $i'$ in the first echelon is assigned to facility $j'$ ($j' = i$; facility $j'$ in the first echelon = client $i$ in the second echelon) as the ${r'}^{th}$ backup facility; AND the demand of client $i$ in the second echelon is assigned to facility $j$ as the $r^{th}$ backup facility<br>$Q_{i',i,j,r,r'} = 0$ ; Otherwise |
| $w_{i',i,r,r'}$ | A binary variable such that:<br>$w_{i',i,r,r'} = 1$ ; if the demand of client $i'$ in the first echelon is assigned to facility $j'$ ($j' = i$; facility $j'$ in first echelon = client $i$ in second echelon) as the ${r'}^{th}$ backup facility; AND demand of client $i$ in the second echelon has $(r-1)$ backup facilities but has no $r^{th}$ backup facility<br>$w_{i',i,r,r'} = 0$ ; Otherwise |
| $\omega_{k,n}$ | A binary variable introduced for the piecewise linearization of the function $\rho_k$ |
| $\varphi_{j,k}$ | A binary variable such that:<br>$\varphi_{j,k} = 1$; if at least one client from regular facility $j$ will assign to mobile facility $k$ as a backup<br>$\varphi_{j,k} = 0$ ; Otherwise |
| $\vartheta_{j,k,n}$ | A binary variable such that:<br>$\vartheta_{j,k,n} = 1$; if regular facility $j$ is assigned as the $n^{th}$ feature of mobile facility $k$<br>$\vartheta_{j,k,n} = 0$ ; Otherwise |
| $\pi_{i,j,k}$ | Continuous variable ($0 \leq \pi_{i,j,k} \leq 1$) indicating whether client $i$ is assigned to regular facility $j$ at the primary backup level ($r = 1$) and also assigned to mobile facility $k$ as a backup |
| $F_{n,k}$ | Continuous variable representing the number of clients assigned to the $n^{th}$ feature of mobile facility $k$, aggregated across all clients and facilities |
| $H_k$ | Continuous variable representing the number of clients simultaneously diverted to mobile facility $k$ |

The mathematical model of VTE-SCNDP based on implicit formulation is as follows:

$$min\, T_1 + T_2 + T_3 \tag{16}$$

$$T_1 = \sum_{l} \sum_{j \leq J_l} s_{l,j} y_{l,j} + \sum_{k \leq K} \delta_k \, \rho_k \tag{17}$$

$$\begin{aligned} T_2 = & \sum_{i \leq I_1} \sum_{j \leq J_1} \sum_{r \leq R_1} c_{1,i,j} \, d_i \, {p_1}^{r-1} \, (1 - p_1) x_{i,j,r} \; + \sum_{k \leq K} c'_{ik} \, p_1 m_{i,k} \; + \\ & \sum_{i \leq I_2 \; and \; j'=i} \; \sum_{j \leq J_2} \sum_{r \leq R_2} \sum_{i' \leq I_1} \sum_{r' \leq R_1} c_{2,i,j} \; (d_{i'} \, \theta_{j'}) {p_1}^{r'-1} \, (1 - p_1) \, {p_2}^{r-1} \, (1 - p_2) \, Q_{i',i,j,r,r'} \; + \end{aligned} \tag{18}$$

$$\begin{aligned} T_3 = & \sum_{r \leq R_1 + 1} {p_1}^{r-1} d_i \, h_{1,i} \, z_{i,r} \\ & + \sum_{i \leq I_2 \; and \; j'=i} \; \sum_{r \leq R_2 + 1} \sum_{i' \leq I_1} \sum_{r' \leq R_1} h_{2,i} \, (d_{i'} \, \theta_{j'}) {p_1}^{r'-1} \, (1 - p_1) \, {p_2}^{r-1} w_{i',i,r,r'} \end{aligned} \tag{19}$$

Equation (16) defines the objective function, which minimizes the total expected cost across both echelons. This includes the fixed costs of opening regular and mobile facilities ($T_1$) in Equation (17), transportation costs ($T_2$) in Equation (18), and penalties incurred for unmet demand ($T_3$) in Equation (19).

$$1 \leq \sum_{j \leq J_l} y_{l,j} \leq \gamma_l \qquad \forall \; l \tag{20}$$

Constraint (20) ensures that the required minimum and maximum number of regular facilities is established in each echelon.

$$\sum_{j \le J_1} x_{i,j,r} + \sum_{r'=1}^{r} z_{i,r'} = 1 \qquad \forall\, i \le I_1;\ \ r = 1 \tag{21}$$

$$\sum_{j \le J_1} x_{i,j,r} + \sum_{k \le K} m_{i,k} + \sum_{r'=1}^{r} z_{i,r'} = 1 \qquad \forall\, i \le I_1;\ \ 2 \le r \le R_1 + 1 \tag{22}$$

Constraint (21) guarantees that each first-echelon client is either assigned to a regular facility at the primary level (i.e., $r = 1$) or declared unmet, in which case a penalty is paid. Additionally, this constraint ensures that mobile facilities are not used under normal operating conditions. Constraint (22) specifies backup-level assignments in the first echelon. It ensures that client $i$ can either be assigned to a facility at backup level $r$ ($r \ge 2$) or the assignment process must have already been halted at a previous level.

$$\sum_{j \le J_1} \sum_{r \ge 2} x_{i,j,r} \le (1 - \sum_{k \le K} m_{i,k})(R_1 - 1) \qquad \forall\, i \le I_1 \tag{23}$$

Constraint (23) specifies that mobile facilities may only serve as backup options at the second level (i.e., $r = 2$) for each client. While this simplifies modeling, it also introduces one of the sources of differences between the IF and SBF.

$$\sum_{j \le J_2} Q_{i',i,j,r,r'} + \sum_{r''=1}^{r} w_{i',i,r'',r'} = x_{i',j',r'} \qquad \forall\, i' \le I_1;\ \ r' \le R_1;\ \ i \le I_2;\ r \le R_2 + 1;\ j' \le J_1 \text{ and } j' = i \tag{24}$$

Constraint (24) manages assignments in the second echelon and has a structure similar to that of Constraints (21) and (22). However, the variables $x_{i',j',r'}$ on its right side, transfer the demand fellow of client $i'$ from the first echelon to the second echelon.

$$\sum_{j \le J_1} x_{i,j,R_1+1} = 0 \qquad \forall\, i \le I_1 \tag{25}$$

$$\sum_{j \le J_2} \sum_{r' \le R_1} Q_{i',i,j,R_2+1,r'} = 0 \qquad \forall\, i' \le I_1;\ \ i \le I_2 \tag{26}$$

Constraints (25) and (26) enforce the payment of a penalty cost when all assigned regular facilities and their backups fail in the first and second echelon respectively.

$$\sum_{r \le R_1} x_{i,j,r} \le y_{1,j} \qquad \forall\, i \le I_1, j \le J_1 \tag{27}$$

$$\sum_{r\le R_2}\sum_{r'\le R_1} Q_{i',i,j,r,r'} \le y_{2,j} \qquad \forall\, i' \le I_1; i \le I_2; j \le J_2 \tag{28}$$

Constraints (27) and (28) prevent assignment to a facility that has not been opened in the first and second echelon, respectively. In addition, these constraints prohibit the assignment of a facility to a client more than once.

$$\sum_{j\le J_1} t_{i,j}\; x_{i,j,r} \le T_{max} \qquad \forall\, i \le I_1;\; r \le R_1 \tag{29}$$

$$\sum_{k\le K} t'_{ik}\; m_{i,k} \le T_{max} \qquad \forall\, i \le I_1 \tag{30}$$

$$\begin{aligned}&\sum_{i\le I_1} d_i\, e_1\left(\sum_{j\le J_1}\sum_{r\le R_1} f_{1,i,j}\, {p_1}^{r-1}\,(1-p_1)x_{i,j,r} + \sum_{k\le K} f'_{ik}\, p_1 m_{i,k}\right) + \\ &\sum_{i\le I_2 | j'=i} e_2\left(\sum_{j\le J_2}\sum_{r\le R_2}\sum_{i'\le I_1}\sum_{r'\le R_1} f_{2,i,j}\left(d_{i'}\beta_{j'}\right){p_1}^{r'-1}\,(1-p_1)\,{p_2}^{r-1}\,(1-p_2)\,Q_{i',i,j,r,r'}\right) \\ &\le E_{max}\end{aligned} \tag{31}$$

Constraints (29) and (30) introduce the travel-time-based agility restriction. Constraint (31) ensures environmental impacts in terms of environmental impact by enforcing a hard cap on the total expected emissions from transportation, which must remain below a given threshold $E_{max}$. Constraints (32) – (38) address the activation probability of mobile facilities.

$$\begin{aligned}\rho_k = \sum_j Pr(A_j) - \sum_{j<j'} Pr(A_j \cap A_{j'}) + \sum_{j<j'<j''} Pr(A_j \cap A_{j'} \cap A_{j''}) - \cdots = \\ 1 - \prod_j \left(1 - Pr(A_j)\right)\end{aligned} \tag{32}$$

Let $Pr(A_j)$ represent the probability that the regular facility $j$ fails under normal conditions while some of its clients are supposed to be served by the mobile facility $k$. Then, Equation (32) expresses the probability of establishing a mobile facility corresponds to the probability that at least one client assigned to it will require service. This is denoted by $\rho_k$ in this equation.

$$\rho_k = 1 - (1-p_1)^{\sum_{j\le J_1}\varphi_{jk}} \tag{33}$$

The probability is expressed as Equation (33) where we assume a uniform failure probability $p_1$, instead of $Pr(A_j)$.

$$\rho_k \ge 1 - (1-p_1)^n - \left(1-\omega_{k,n}\right).G \qquad \forall\, k \le K;\; 1 \le n \le \gamma_1 \tag{34}$$

$$\sum_{1\le n\le \gamma_1} \omega_{k,n} = 1 \qquad \forall\, k \le K \qquad (35)$$

$$(n - \sum_{j\le J_1} \varphi_{j,k}) + (\omega_{k,n} - 1).G \le 0 \qquad \forall\, k \le K; 1 \le n \le \gamma_1 \qquad (36)$$

$$(\sum_{j\le J_1} \varphi_{j,k} - n) + (\omega_{k,n} - 1).G \le 0 \qquad \forall\, k \le K;\ 1 \le n \le \gamma_1 \qquad (37)$$

$$\sum_{i\le I_1} \pi_{i,j,k} - \varphi_{j,k}\, G \le 0 \qquad \forall\, j \le J_1;\ k \le K \qquad (38)$$

Because Constraint (33) is nonlinear, it is reformulated and linearized in Constraint (34) – (38) using a piecewise method combined with Big-M.

$$Pr(H_k \le MC_k) \ge\ 95\% \qquad \forall\, k \le K \qquad (39)$$

Constraint (39) imposes the probabilistic requirement that ensures a high level of service reliability under uncertainty. Unlike our earlier SBF that imposed a strict deterministic limit $MC_k$, relying solely on $MC_k$ assumes a worst-case situation. This often makes the model unnecessarily restrictive and can reduce solution quality. Here, we instead adopt a probabilistic service level, with 95% confidence, no more than $MC_k$ clients should require simultaneous service from a mobile facility $k$.

Since partial assignment is not allowed in the problem, decomposing the second-echelon allocation variables for client $i$ (i.e. $w_{i',i,r,r'}$ and $Q_{i',i,j,r,r'}$) into $I_1$ independent variables due to the additional index $i' \le I_1$ may appear to permit partial allocation of client demand. However, Theorem 1 guarantees that all client demand is fully assigned in the second echelon, and therefore partial allocation cannot occur.

**Theorem1.** *In the optimal solution, if* $Q_{i',i,j,r,r'} = 1$ *and* $Q_{i'',i,j',r,r''} = 1$ *then* $j = j'$*; that is, partial assignment of a client's demand across multiple facilities does not occur.*

**Proof.** We prove the result by contradiction. Suppose that, in an optimal solution, there exists a client $i$ such that $Q_{i',i,j,r,r'} = 1$ *and* $Q_{i'',i,j',r,r''} = 1$ with $j \ne j'$, implying that the demand of client $i$ is partially assigned to two different regular facilities in the second echelon. Without loss of generality, assume that $c_{2,i,j} < c_{2,i,j'}$.

Consider reassigning the portion of demand originally served by facility $j'$ to facility $j$, while keeping all other decision variables unchanged. This reassignment is feasible since both facilities serve the same client under identical scenario indices. Moreover, the reassignment preserves feasibility with respect to Constraint (31) because the transportation cost $c_{l,i,j}$ is a function of the distance $f_{l,i,j}$. Consequently, $c_{2,i,j} < c_{2,i,j'}$ implies $f_{2,i,j} < f_{2,i,j'}$, and swapping the assignment from $j'$ to $j$ reduces the left-hand side of Constraint (31).

$$\Delta T_2 = (c_{2,i,j'} - c_{2,i,j})\big(d_{i'}\,\theta_i\big)p_1^{\,r''-1}\left(1-p_1\right)p_2^{\,r-1}\left(1-p_2\right) \tag{40}$$

The resulting change in the second-echelon transportation cost $T_2$ is calculated in Equation (40), which is strictly positive due to $c_{2,i,j} < c_{2,i,j'}$. Therefore, the reassignment strictly reduces the total expected cost, contradicting the optimality of the original solution. Hence, $j = j'$, and partial assignment of a client's demand across multiple facilities cannot occur in an optimal solution. □

## 5. Incorporating machine learning into IF

Directly enforcing the chance constraint $Pr(H_k \leq MC_k) \geq 95\%$ is computationally intractable, since the random variable $H_k$, depends on how clients are distributed among regular facilities and how these facilities fail simultaneously. In particular, $H_k$ represents the number of clients that may be simultaneously redirected to a mobile facility due to disruptions, which is a sum of independent Bernoulli failure events, whose probabilities are directly decision-dependent (i.e., they depend on how the clients who are supposed to be reassigned to the mobile facility are distributed across opened regular facilities). As a result, even evaluating $Pr(H_k \leq MC_k)$ for a given solution requires enumerating a large number of joint failure combinations, and no closed-form expression for its distribution exists.

Classic approaches for handling such chance constraints face significant limitations in this setting. Scenario-based methods would require explicitly enumerating a very large number of disruption scenarios to accurately capture low-probability but high-impact events, leading to exponential growth in model size and computational burden (Prékopa, 2013). On the other hand, analytical approximations such as Chebyshev-type bounds or CVaR reformulations rely on worst-case or distribution-free estimates, which are known to be overly conservative for discrete and highly skewed distributions (Nemirovski & Shapiro, 2007), resulting in suboptimal and overly restrictive solutions. To address these issues, we replace the explicit chance constraint with a learned linear cut. This cut separates feasible assignment patterns (those satisfying the 95% capacity requirement) from infeasible ones, providing an efficient data-driven approximation.

### 5.1. Generating training data set

To learn the linear cut, we construct a labeled dataset that captures how different client assignment patterns affect the probability of exceeding the mobile facility capacity. For each assignment, we determine the number of clients that would need to be reassigned to a mobile facility following failures of their corresponding regular facilities. These values are derived directly from how clients are distributed among the opened regular facilities.

To eliminate permutation redundancy and ensure a consistent representation, these values are arranged in a non-increasing order. This step is essential, as different assignments that are equivalent up to permutation lead to identical probability profile, and sorting allows the learning model to recognize them as the same pattern.

For each resulting pattern, we compute the exact value of $Pr(H_k \leq MC_k)$ by enumerating all relevant failure combinations. Based on a predefined confidence level (e.g., 95%), patterns are labeled as feasible or infeasible. The trained model therefore learns the relationship between the structure of client assignments and the corresponding overflow risk.

The key point is that the trained classifier does not depend directly on economic parameters such as demand levels, transportation costs, or distances. Instead, it learns the relationship between client assignment patterns across opened regular facilities, which is directly related to the failure probability of facilities in first echelon ($p_1$), maximum number of regular facilities that must be established in echelon 1 ($\gamma_1$) and mobile facility capacity threshold ($MC_k$). In other words, by changing $\gamma_1$ the distribution of the clients who are supposed to be reassigned to the mobile facility changes, which in turn affects the probability of each Bernoulli failure event. Also, changing $MC_k$ and $p_1$ directly affects $Pr(H_k \leq MC_k)$. Therefore, the trained model generalizes well across different problem instances, as long as $\gamma_1$, $p_1$, and the $MC_k$ remain unchanged. This makes the approach reusable across a wide range of demand, cost, and other parameter settings. However, if these key structural parameters are modified, retraining becomes necessary. Table 2 presents a sample subset of a training dataset, including selected rows and feature columns that illustrate its structure, where $\gamma_1$, $p_1$, and the $MC_k$ are equal to 5, 0.15, and 3 respectively. In this table, each row corresponds to a client assignment pattern across regular facilities. The "Regular Facility" columns indicate how clients who are supposed to be reassigned to a mobile facility at the second backup level ($r = 2$) are assigned to the opened facilities. The "ML Features" are same data, arranged in non-increasing order to eliminate permutation redundancy, while these values must be normalized prior to training. The probability $Pr(H_k \leq MC_k) = \sum_{H_k=0}^{3} Pr(H_k)$ is computed for each pattern, and the corresponding data label indicates whether the assignment satisfies the required service level 0.95 (label 1) or not (label 0).

**Table 2.** Sample of the training dataset illustrating the structure of input features and corresponding labels

| **Regular Facility No.** | | | | | **ML Features No.** | | | | | $Pr(H_k \leq MC_k)$ | **Data label** |
|---|---|---|---|---|---|---|---|---|---|---|---|
| **1** | **2** | **3** | **4** | **5** | **1** | **2** | **3** | **4** | **5** | | |
| 0 | 2 | 0 | 0 | 3 | 3 | 2 | 0 | 0 | 0 | 0.9775 | 1 |
| 1 | 2 | 0 | 1 | 0 | 2 | 1 | 1 | 0 | 0 | 0.9966 | 1 |
| 1 | 0 | 1 | 3 | 1 | 3 | 1 | 1 | 1 | 0 | 0.9421 | 0 |
| 3 | 0 | 1 | 1 | 0 | 3 | 1 | 1 | 0 | 0 | 0.9583 | 1 |
| 1 | 1 | 4 | 1 | 0 | 4 | 1 | 1 | 1 | 0 | 0.85 | 0 |
| 1 | 2 | 1 | 1 | 2 | 2 | 2 | 1 | 1 | 1 | 0.9620 | 1 |

## 5.2. Training and implementation

We train several linear machine learning models on the data set, including Logistic Regression, L1-regularized Logistic Regression, a linear SVM via SGD Classifier, a Perceptron, and Logistic Regression with $C = 0.1$ (stronger regularization). We select the best model based on held-out accuracy. To improve numerical stability and learning performance, all feature vectors are normalized prior to training using standardization.

$$x_{norm,i} = (x_i - \mu_i)/\sigma_i \tag{41}$$

Specifically, each feature is transformed as Equation (41) where $\mu_i$ and $\sigma_i$ denote the mean and standard deviation of feature $i$, respectively. The machine learning models are trained on these normalized features. Therefore, the learned classifier takes the form $z = b_{norm} + \sum_i w_{norm,i} \cdot x_{norm,i}$.

However, since the optimization model operates on the original (unnormalized) decision variables, the learned coefficients must be transformed back into the original feature space. Substituting the normalization expression into the classifier yields $z = b_{raw} + \sum_i w_{raw,i} \cdot x_i$.

$$w_{raw,i} = \frac{w_{norm,i}}{\sigma_i} \quad \text{and} \quad b_{raw} = b_{norm} - \sum_i \frac{w_{norm,i} \cdot \mu_i}{\sigma_i} \tag{42}$$

Equation (42) calculates $w_{raw,i}$ and $b_{raw}$ accordingly. This transformation ensures that the learned classifier can be directly embedded into the MILP formulation without altering its predictive behavior.:

$$\text{Z} = \text{intercept} + \sum_n Z_{coef}[n] \times \text{Features}[n] \geq 0. \tag{43}$$

After converting the learned coefficients back to the scale of the original (unnormalized) features, one linear constraint per mobile facility is added into the MILP as represented in Equation (43). The feature variables are arranged in non-increasing order so that they match the classifier's feature ordering. Enforcing $Z \geq 0$ corresponds to predicting Class 1, meaning that the assignment is expected to satisfy the $\geq 95\%$ capacity-protection requirement. From a geometric perspective, the learned classifier defines a hyperplane in the feature space that separates feasible and infeasible assignment patterns. This learned hyperplane enforces that only patterns lying on the feasible side of this boundary are allowed in the optimization model.

To illustrate the transition from the probabilistic Constraint (39) to the learned linear cut (44), consider a representative dataset with $\gamma_1$, $p_1$, and the $MC_k$ equal to 5, 0.15, and 3 respectively. For this dataset, after finding the best-performing classifier and transforming the learned coefficients back to the original (unnormalized) feature space, the resulting linear classifier is:

$$z = 11.338883 - 3.166690x_1 - 0.013761x_2 - 1.581756x_3 - 1.012822x_4 + 0.354179x_5$$

where $x_1, \ldots, x_5$ denote the sorted numbers of clients potentially reassigned to the mobile facility from the opened regular facilities, arranged in non-increasing order. A pattern is classified as feasible whenever $z \geq 0$, which corresponds to the positive class learned from the training data. Consider the assignment pattern (1,0,1,3,1). After sorting in non-increasing order, the feature vector becomes (3,1,1,1,0). Substituting this pattern into the learned linear cut gives:

$$z = 11.338883 - 3.166690(3) - 0.013761(1) - 1.581756(1) - 1.012822(1) + 0.354179(0) = -0.769526$$

Since $z < 0$, this pattern is classified as infeasible by the learned constraint. We can verify this result by directly evaluating the original probabilistic condition. Under this pattern, the mobile facility may need to serve 3, 1, 1, 1, and 0 clients depending on which regular facilities fail. Enumerating all failure combinations with independent facility failure probability 0.15 gives:

$$Pr(H_k \leq 3) = 0.94211875$$

Since $0.94211875 < 0.95$, the pattern violates the required 95% service level and is therefore truly infeasible.

Now consider the assignment pattern (1,2,1,1,2). After sorting in non-increasing order, the feature vector becomes (2,2,1,1,1). Substituting this pattern into the learned linear cut yields:

$$z = 11.338883 - 3.166690(2) - 0.013761(2) - 1.581756(1) - 1.012822(1) + 0.354179(1) = 2.737582$$

Since $z \geq 0$, the pattern is classified as feasible by the learned constraint. Direct evaluation of the original chance constraint confirms this prediction:

$$Pr(H_k \leq 3) = 0.96200875$$

Because $0.96200875 \geq 0.95$, this pattern satisfies the required service level and is therefore truly feasible.

### 5.3. Corresponding constraints

For applying the linear classifier to ensure that the client distribution pattern associated with mobile facility $k$ (represented by variables $\mathcal{F}_{n,k}$) meets the 95% probability threshold, we reinterpret Constraint (39) through the deterministic reformulation given in Constraints (44) – (50).

$$\beta_{n,k} + \sum_{1 \leq n \leq \gamma_1} \alpha_{n,k} \mathcal{F}_{n,k} \geq 0 \qquad \forall\, k \leq K \qquad (44)$$

Constraint (44) is the learned linear cut that ensures only assignment patterns predicted to meet the capacity requirement with at least 95% probability is accepted.

$$\sum_{j\le J_1} \vartheta_{j,k,n} \le 1 \qquad \forall\, k \le K; 1 \le n \le \gamma_1 \qquad (45)$$

$$\sum_{1\le n\le\gamma_1} \vartheta_{j,k,n} \le y_{1,j} \qquad \forall j \le J_1;\ k \le K \qquad (46)$$

$$\sum_{1\le n\le\gamma_1} \sum_{j\le J_1} \vartheta_{j,k,n} = \sum_{j\le J_1} y_{1,j} \qquad \forall\, k \le K \qquad (47)$$

Constraints (45) – (47) define the input features for the linear classifier. These constraints establish a one-to-one mapping between open facilities and classifier features for each mobile facility $k$.

$$\pi_{i,j,k} \ge x_{i,j,1} - (1 - m_{i,k}) \qquad \forall\, i \le I_1;\ j \le J_1;\ k \le K \qquad (48)$$

Constraint (48) links variable $\pi_{i,j,k}$ to the primary facility assignment ($x_{i,j,1}$) and mobile assignment ($m_{i,k}$), ensuring $\pi_{i,j,k} = 1$ only when client $i$ is assigned to both regular facility $j$ and mobile facility $k$ at levels 1 and 2 respectively.

$$\mathcal{F}_{n,k} \ge \sum_{i\le I_1} \pi_{i,j,k} - (1 - \vartheta_{j,k,n})G \qquad \forall j \le J_1;\ k \le K;\ 1 \le n \le \gamma_1 \qquad (49)$$

$$\mathcal{F}_{n,k} \le \sum_{i\le I_1} \pi_{i,j,k} + (1 - \vartheta_{j,k,n})G \qquad \forall j \le J_1;\ k \le K;\ 1 \le n \le \gamma_1 \qquad (50)$$

Constraints (49) and (50) set variable $\mathcal{F}_{n,k}$ to be at least equal to the total number of clients transferred from regular facility $j$ to mobile facility $k$, where regular facility $j$ corresponds to $n^{th}$ feature. These constraints ensure that any feasible distribution pattern remains feasible with respect to Constraint (44).

$$\mathcal{F}_{n,k} \ge \mathcal{F}_{n+1,k} \qquad \forall\, k \le K;\ 1 \le n \le \gamma_1 - 1 \qquad (51)$$

Constraint (51) enforces a non-increasing sequence for $\mathcal{F}_{n,k}$, representing the value of the $n^{th}$ feature for the mobile facilities $k$. This ordering constraint eliminates permutation symmetry in the assignment representation, ensuring that the input pattern matches the classifier's training format and improving prediction accuracy.

## 6. Fix-and-relax heuristic

Before explaining the details of the proposed fix-and-relax heuristic, it is important to indicate the reason for choosing this heuristic. Although metaheuristic algorithms have been widely used in supply chain network design, their application to the current problem poses several challenges. The existence of probabilistic service constraints and also cross-echelon dependencies should be preserved to ensure feasibility in all interdependent decision layers. In metaheuristic algorithms, these constraints are generally handled using a penalty mechanism, which does not guarantee a feasible solution. Therefore, the solution's quality may be unstable. The proposed solution approach, in contrast, specifically keeps feasibility via integrating machine-learning–derived linear cuts into a mathematical programming framework. Moreover, the fix-and-relax heuristic utilizes the problem's structural properties to decompose the decision space. It then applies exact solvers to efficiently solve the resulting subproblems. Finally, not only is extensive parameter calibration unnecessary, but there is also an applicable and robust solution approach.

The fix-and-relax (FR) heuristic is a constructive heuristic that builds a solution step by step by dividing the problem into segments. In each iteration, the key integer variables of the current segment are optimized, the past segments are fixed to preserve feasibility, and the future segments are relaxed to simplify the model. By moving forward and gradually fixing decisions, F&R efficiently constructs a high-quality solution for large mixed-integer problems that are otherwise too hard to solve directly. In this paper, we develop two versions of the fix-and-relax heuristic: the Backup-Based Fix-and-Relax (BBFR) and the Echelon-Based Fix-and-Relax (EBFR). The detailed procedures for each version are presented below.

### 6.1 Backup-based fix-and-relax

In the proposed heuristics, the main model is decomposed into $R_1 + R_2$ sub-problems, with each sub-problem corresponding to a specific backup level in the original model. This procedure begins with backup level 1 in the first echelon and proceeds through each level up to the final backup level in the second echelon ($R_2$), completing a total of $R_1 + R_2$ iterations.

Because demand flows aggregate from the lower echelon to the upper echelon in our model, the first-echelon decisions have greater strategic influence on the overall network configuration. For this reason, the heuristic begins with the first echelon. Moreover, the solution is developed from the initial backup levels toward the later ones, since early-level decisions have a much stronger impact on system performance. In contrast, the cost of services and penalties in higher backup levels is relatively small because the probability of reaching these levels decreases rapidly as facility-failure probabilities are multiplied sequentially.

In each iteration, one of three strategies can be applied: a **freezing strategy**, a **full-model strategy**, or a **relaxation strategy**. In the first iteration, all variables and constraints associated with backup level 1 are kept in their original form. For the remaining levels, from level 2 in the first echelon up to level $R_2$ in the second echelon, the relaxation strategy is used to simplify the model by relaxing the binary variables. Similarly, in the $r^{th}$ iteration, the same three strategies are applied in a structured way: the variables for level $r$ are treated exactly, previously solved levels are frozen, and the variables for all later levels remain relaxed. Note that the $y_{l,j}$ in all iterations are exempted from relaxation or freezing strategies. The details of the strategies are as follows:

### *Freezing strategy*

In the freezing strategy for the $r^{th}$ iteration, previously optimized binary variables are fixed to their values from iteration $r-1$.

- If $r \leq R_1 + 1$, then the binary variables $x_{i,j,r-1}$ and $z_{i,r-1}$ (associated with the first echelon) are fixed to their optimal values from the $(r-1)^{th}$ iteration.
- If $R_1 + 1 < r \leq R_1 + R_2$, then the binary variables $Q_{i',i,j,r-R_1-1,r'}$ and $w_{i',i,r-R_1-1,r'}$, (associated with the second echelon) are fixed based on the optimal solution of the $(r-1)^{th}$ iteration.

***Note.*** All binary variables related to capacity planning and the establishment probability of mobile facilities ($m_{i,k}$, $\omega_{k,n}$, $\varphi_{j,k}$, $\vartheta_{j,k,n}$) are frozen after the second iteration ($r \geq 3$).

### *Solving the full-model*

In the full-model strategy of the $r^{th}$ iteration, if ($r \leq R_1$), the binary variables associated with the current backup level are solved exactly (without relaxation or fixing). Specifically:

- If $r \leq R_1$, then the binary variables for the $r^{th}$ backup level of the first echelon ($x_{i,j,r}$ and $z_{i,r}$) are fully optimized.
- If $R_1 < r \leq R_1 + R_2$, then the binary variables related to the $(r-R_1)^{th}$ backup level of second echelon ($Q_{i',i,j,r-R_1,r'}$ and $w_{i',i,r-R_1,r'}$), are fully optimized.

***Note.*** All binary variables related to capacity planning and the establishment probability of mobile facilities ($m_{i,k}$ , $\omega_{k,n}$ , $\varphi_{j,k}$ , $\vartheta_{j,k,n}$) are optimized during the second iteration ($r = 2$).

***Relaxation strategy***

In the relaxation strategy of the $r^{th}$ iteration:

- If $r \leq R_1 - 1$, the binary variables corresponding to backup levels $r + 1$ through $R_1$ in the first echelon ($x_{i,j,r'}$ and $z_{i,r'} \mid r' = r + 1, \dots, R_1$) are relaxed. In addition, all binary variables associated with backup levels 1 to $R_2$ in the second echelon ($Q_{i',i,j,r',r''}$ and $w_{i',i,r',r''} \mid r' = 1, \dots, R_2$) are relaxed.
- If $R_1 \leq r \leq R_1 + R_2$, the binary variables corresponding to backup levels $r - R_1 + 1$ through $R_2$ in the second echelon ($Q_{i',i,j,r',r''}$ and $w_{i',i,r',r''} \mid r' = r - R_1 + 1, \dots, R_2$) are relaxed.

***Note.*** All binary variables related to capacity planning and the establishment probability of mobile facilities ($m_{i,k}$ , $\omega_{k,n}$ , $\varphi_{j,k}$ , $\vartheta_{j,k,n}$) are relaxed only during the first iteration ($r = 1$).

We have presented an algorithmic description of the proposed BBFR heuristic in Algorithm 1.

**Algorithm 1.** The proposed BBFR heuristic (The first versions of the fix-and-relax heuristic)

**Inputs:** IF model parameters, cost/distance matrices, failure probabilities, and operational limits.
**Outputs:** Facility opening decisions, backup assignments, mobile facility plans, and total cost.
**For** $l = 1 \; to \; 2$
    **For** $r = 1 \; to \; R_l$
        **If** $l = 1$ **then**
            $x_{i,j,r-1}$ and $z_{i,r-1}$ and ($m_{k,j}$ , $\omega_{k,n}$ , $\varphi_{j,k}$ , and $\vartheta_{j,k,n}$ if $r > 2$) are fixed into their values in previous iteration
            ($x_{i,j,r'}$ and $z_{i,r'} \mid r' = r + 1 \; to \; R_1$) and ($Q_{i',i,j,r',r''}$ and $w_{i',i,r',r''} \mid r' = 1 \; to \; R_2$) and ($m_{k,j}$ , $\omega_{k,n}$ , $\varphi_{j,k}$ , and $\vartheta_{j,k,n}$ if $r = 1$) are relaxed
            $x_{i,j,r}$ and $z_{i,r}$ and ($m_{k,j}$ , $\omega_{k,n}$ , $\varphi_{j,k}$ , and $\vartheta_{j,k,n}$ if $r = 2$) are considered as the binary variable
        **Else**
            **If** $r = 1$ **then**
                $x_{i,j,R_1}$ and $z_{i,R_1}$ are fixed based on optimum value of previous iteration
            **Else**
                $Q_{i',i,j,r-1,r'}$ and $w_{i',i,r-1,r'}$ are fixed based on optimum value of previous iteration
            ($Q_{i',i,j,r',r''}$ and $w_{i',i,r',r''} \mid r' = r + 1, \dots, R_2$) are relaxed
            $Q_{i',i,j,r,r'}$ and $w_{i',i,r,r'}$ are considered as the binary variable
        Solve the new MIP model
        If all optimal variables are binary, then stop the algorithm, and the final solution is obtained
**Return** Outputs;

## 6.2 Echelon-based fix-and-relax

In the echelon-based fix-and-relax heuristic, the original IF model is divided into two sub-problems. Each sub-problem is equivalent to a single echelon in the original model. In the first iteration, the variables and constraints related to the first echelon are considered as the **full-model strategy** (without changes), and the binary variables related to the second echelon, including ($Q_{i',i,j,r,r'}$ and $w_{i',i,r,r'}$) are relaxed. In the second iteration, the binary variables related to the first echelon ($y_{1,j}, x_{i,j,r}, z_{i,r}, m_{i,k}, \omega_{k,n}, \varphi_{j,k}$, and $\vartheta_{j,k,n}$) are fixed based on the optimum value of the first iteration, and the model related to the second echelon is solved completely (no changes include relaxation or fixed). We have presented an algorithmic description of the proposed EBFR heuristic in Algorithm 2.

**Algorithm 2.** The proposed EBFR heuristic (The second version of the fix-and-relax heuristic)

```
Inputs: IF model parameters, cost/distance matrices, failure probabilities, and operational limits.
Outputs: First-echelon decisions, second-echelon assignments, and total cost.
For  l = 1 to 2
    If  l = 1  then
        (Q_{i',i,j,r,r'} and w_{i',i,r,r'} | r = 1 to R_2) are relaxed
    Else
        y_{1,j}, x_{i,j,r}, z_{i,r}, m_{k,j}, ω_{k,n}, φ_{j,k}, and ϑ_{j,k,n} are fixed based on the optimum value of the first iteration
        (Q_{i',i,j,r,r'} and w_{i',i,r,r'} | r = 1 to R_2) are considered as the binary variable
    Solve the new MIP model
    If all optimal variables are binary then stop the algorithm and the final solution is obtained
Return Outputs;
```

### 6.3. Local search

To further enhance the solutions generated by the proposed fix-and-relax algorithms, we apply a post-processing local search. This local search improves solution quality by fixing the values of selected decision variables, releasing others, and re-optimizing the model accordingly. In this local search, in addition to the dependent variables $\omega_{k,n}$, $\varphi_{j,k}$, and $\vartheta_{j,k,n}$, all of the decision variables for special clients $i$ in the first echelon ($x_{i,j,r}$, $z_{i,r}$ and $m_{i,k}$) and second echelon ($Q_{i,i',j,r,r'}$ and $w_{i,i',r,r'}$) are released and then resolved again.

A key component of this local search is deciding which clients, and how many of them, should have their variables released. Proper selection enables significant solution improvements within a short computational time. Releasing more decision variables can improve solution quality, but selecting too many clients greatly increases computational time. Conversely, releasing only a few clients, or choosing the wrong ones, yields limited benefits.

The proportion of released clients is determined based on computational experiments. In particular, we tested several candidate values (e.g., 10%, 15%, 20%, and higher) and found that releasing approximately 15% of clients with the highest demand provides a good balance between solution quality and computational efficiency. We note that this value is not necessarily optimal, but rather represents a practical approximation based on empirical observations. These experiments were conducted primarily on large-scale instances, where the local search procedure plays a more critical role. In smaller instances, the initial solutions obtained by the fix-and-relax algorithm are already of high quality, and thus the impact of local search is limited.

Although it is possible to tune this parameter for different instance sizes, we intentionally adopt a fixed percentage to preserve the generality and practical applicability of the approach. Since this ratio is primarily derived from experiments on large-scale instances, decision-makers may consider using larger ratios for smaller instances to further explore potential improvements in solution quality without significant computational burden.

Finally, when multiple clients have identical demand values and cannot all be included within the selected percentage, a random selection is applied among them. This ensures simplicity and avoids bias in the selection process.

## 7. Computational results

In this section, the credibility of the presented mathematical models and the performance and effectiveness of the presented approximation algorithms are evaluated and compared with each other and with other mathematical models. To run the developed solution methods in each experiment, the IBM CPLEX Optimization Studio 20.1.0 solver is employed, and its related DOcplex.MP library is imported into the Python 3.8 programming language. All the related scripts are executed on a computer with an Intel (R) Core (TM) i7-8565U processor and 16 GB of RAM.

$$RPD1_A = \frac{(objectve\ function)_A - Minimum(objectve\ functions)}{Minimum(objectve\ functions)} \tag{52}$$

$$RPD2_A = \frac{(objectve\ function)_A - Average(objectve\ functions)}{Average(objectve\ functions)} \tag{53}$$

To compare the performance of the solution methods and demonstrate their effectiveness, in addition to GAP% (which measures the gap from the lower bound), two Relative Percentage Deviation metrics, RPD1 and RPD2, are also used. These performance measures are calculated based on the deviation of solutions from the best solution and the average of solutions achieved by mathematical models, BBFR, EBFR, and SAA. The RPD criteria for solution method A are calculated in Equations (52) and (53). Note that the index $A$ denotes a solution method.

### 7.1 Experimental Setup and Instance Generation

This section describes the testing of the proposed algorithms on 20 data sets that are generated randomly. The data set consists of 28 to 498 nodes for small to large instances.

In each case, demands $d_i$ in the first echelon are drawn from a normal distribution with ($\mu = 100$ ; $\delta = 30$) and rounded to the nearest integer. Facility fixed costs are sampled uniformly and rounded: $s_{1,j} \sim U[30000{,}80000]$, $s_{2,j} \sim U[120000{,}200000]$, and mobile deployment costs $\delta_k \sim U[5000{,}15000]$. Penalty costs $h_{1,i}$ and $h_{2,i}$ are drawn from $U[500{,}1500]$ and $U[1000{,}3000]$ respectively and rounded to the nearest integer. In addition, the conversion coefficients $\theta_j$ are drawn from $U[1{,}2]$.

Nodes are placed uniformly at random in a square $[0, \partial]^2$, where $\partial = max\ (5, 10\sqrt{I_1 + J_1 + J_2}$. Transportation costs $c_{l,i,j}$ and $c'_{i,k}$ are set equal to the Euclidean distance between $i$ and $j$. Travel times $t_{l,i,j}$ and $t'_{i,k}$ divide distances by a speed parameter ($travel_speed \sim U[0.5{,}1]$). A global travel-time bound is set to $T_{max} = (Tmax_factor\ .\ \partial)/travel_speed$ with $Tmax_factor = 0.5$. Each mobile site has a deterministic nominal capacity parameter $MC_k \in [2{,}3{,}4{,}5]$ dependent on the instance size. Emissions use rate ($e_l = 1 | l = 1{,}2$), and a cap $E_{max}$ defined as 70% of a conservative bound combining first- and second-echelon worst-case travel. The full failure probability of facilities in the first echelon is equal to 0.15, and in the second echelon is equal to 0.12. The scenarios of each instance are generated simply by considering all possible combinations of active and inactive facilities. The probability of each scenario is computed by multiplying the probabilities of facilities according to their situation in the scenario (active or inactive).

Ultimately, each instance is labeled with $(a; b; c; d)$ , where $a$ indicates the number of clients in the first echelon; $b$ indicates the number of potential sites in the first echelon (potential clients in the second echelon), $c$ indicates the number of potential sites in the second echelon, and $d$ indicates the number of potential sites for the mobile facilities. The full set of instances is publicly accessible at: https://zenodo.org/records/19660479. The size of these parameters influences both the solution quality and the efficiency of the proposed procedures. In order to find the best trade-off between the algorithm speed and solution quality, the runtime is limited to 3600 seconds.

### 7.2 Numerical analysis

In this subsection, we present the results of the numerical examples to show the validity of the proposed SBF and IF and the efficiency of the presented solution methods.

Table 3 shows a comparison of the results obtained with two mathematical models: the IF and SBF. The results indicate a clear performance difference between the two formulations. The scenario-based formulation (SBF) is capable of finding feasible solutions only for small instances and fails to produce any feasible solution for problems larger than sp(80;8;2;10). In contrast, the implicit formulation (IF) demonstrates strong performance on small- and medium-sized instances, achieving an average RPD1 of 5.5%, which highlights its superiority over SBF in these cases. However, this advantage diminishes as the problem size increases. For large-scale instances, the performance of IF deteriorates significantly, with the average RPD1 rising to 418%, indicating a substantial loss in solution quality. This behavior underscores the necessity of developing more efficient solution algorithms to effectively handle large-scale instances. However, considering the average difference of less than 5% between optimal solutions of the SBF and IF, and the inefficiency of the SBF for large-sized problems, we can propose IF as a suitable alternative to SBF and an efficient model for the VTE-SCNDP.

In Table 3, we observe that the IF model yields objective values that are slightly higher than those produced by the detailed SBF. This behavior is expected, as the implicit formulation is structurally more restrictive (see Figure 2). However, the discrepancy does not arise solely from this structural difference and requires further clarification. In particular, an additional source of discrepancy stems from the probabilistic enforcement of the mobile-facility capacity constraint in the IF model. Specifically, the constraint is satisfied with at least 95% confidence, but violations occurring in the remaining <5% of low-probability failure events are not penalized in the objective function. As a result, the IF model may underestimate the true expected system cost. In contrast, the SBF explicitly enumerates all possible failure scenarios, and, therefore, must eliminate capacity violations in every scenario. To correct this downward bias, we introduce a post-optimality adjustment to the IF's objective function (Modified objective function). After solving the IF, we evaluate the expected cost of mobile-capacity violations implied by the chosen solution. We have presented an algorithmic description of the post-optimality adjustment of the IF in Algorithm 3.

**Algorithm 3: Post-optimality objective correction for Implicit Formulation (IF)**

**Inputs:** IF solution and parameters.
**Outputs:** Adjusted penalty and corrected objective.
*additional_penalty* ← **0**
**For** $k$ *in* $K \mid \sum_{i \leq I_1} m_{i,k} \geq 1$
    *facilities_set* ← [$j$; $j$ in $J_1$ **if** $\sum_{i \leq I_1} \pi_{i,j,k} \geq 1$]
    *total_facilities* ← **Length** (*facilities_set*)
    $S$ ← set of 2^{*total_facilities*} scenarios; all combinations of active/inactive facilities in the *facilities_set*
        **For** each scenario $s$ in $S$
            *clients_set* ← **[ ] ;** *n_inactive* ← **0**
            **For** $j$ in *facilities_set*
                **If** $j$ is inactive in scenario $s$ **Then**
                    *n_inactive* ← *n_inactive* **+ 1**
                    *clients_set* ← *clients_set* ∪ **[**$i$; $i \leq I_1$ **if** $\pi_{i,j,k} = 1$**]**
            **If Length** (*clients_set*) $> MC_k$ **Then**
                $P_s$ ← $p_1$^{*n_inactive*}× $(1 - p_1)$ ^{*total_facilities* − *n_inactive*}
                **Sort** *clients_set* by $h_{1,i}$ ascending
                **For** $i$ in *clients_set* [1**:** **Length** (*clients_set*) $- MC_k$]
                    *additional_penalty* ← *additional_penalty* + $(h_{1,i} \times d_i \times P_s)$
*Modified objective function* ← *IF objective function* + *additional_penalty*
**Return** Outputs;

Table 4 shows a comparison of the results obtained by the proposed fix-and-relax heuristics: backup-based fix-and-relax (BBFR) and echelon-based fix-and-relax (EBFR). It shows that both proposed heuristic methods solve the IF very efficiently. BBFR yielded an acceptable performance by consuming much less time than CPLEX Solver and an average RPD1 of less than 2.2%. In addition, EBFR obtained the optimal or near optimal solution in all cases. The BBFR and EBFR reduce CPU time by approximately 4000 seconds on average across the 20 solved instances.

Because the full SBF becomes computationally intractable even for small problem sizes, an SAA procedure was developed to generate high-quality approximations more efficiently (see **Appendix A** for details). Table 5 reports the results obtained using the SAA method, showing that it enables the SBF to be applied to larger-sized instances. On the other hand, the average RPD1 equal to 4.3% for sp(18;5;2;3) to sp(160;15;4;35) instances reflects the quality of the final solutions of SAA. This algorithm could reduce the total solution time of sp(18;5;2;3) to sp(120;12;3;15) instances from >21614 seconds to 8796 seconds. We recommend the use of the SAA method for the SBF of the VTE-SCNDP. However, it should not be regarded as a general rule. In more detail, the results of our experiment in Tables 4 and 5 demonstrate that SAA is especially suitable for medium-sized problems solved by the SBF, whereas BBFR and EBFR heuristics are preferable for large-sized problems solved by the IF.

**Table 3.** Comparison of the results obtained by the IF and SBF

| Sample problems | SBF (CPLEX Solver) | | | | | | | | IF (CPLEX Solver) | | | | | | | | |
|---|---|---|---|---|---|---|---|---|---|---|---|---|---|---|---|---|---|
| | Time (s) | Objective function | T1 | T2 | T3 | Lower Bound | Gap% | RPD1% | Time (s) | Objective function | Modified objective function | T1 | T2 | T3 | Lower Bound | Gap% | RPD1% |
| sp(18;5;2;3) | 2.5 | 1006942.1 | 332625.2 | 80974.6 | 593342.3 | 1006942.1 | 0.0% | 0.0% | 0.3 | 1085419.9 | 1147556.4 | 327936.5 | 80260.8 | 739359.1 | 1085419.9 | 0.0% | 7.8% |
| sp(30;5;2;5) | 11.4 | 1569669.2 | 330947.9 | 136934.9 | 1101786.4 | 1569669.2 | 0.0% | 0.0% | 33.7 | 1573106.5 | 1638363.7 | 320568.2 | 132147.6 | 1185648 | 1573106.5 | 0.0% | 0.2% |
| sp(50;8;2;5) | >3600 | 1883116.6 | 328647.2 | 322653 | 1231816.4 | 1881677.0 | 0.1% | 0.0% | >3600 | 1985395.7 | 2005483.3 | 270230.2 | 313958.7 | 1421294.5 | 1958614.2 | 1.4% | 5.4% |
| sp(80;8;2;10) | >3600 | 3946716.4 | 522101.5 | 665142.7 | 2759472.2 | 1171373.6 | 236.9% | 42.4% | >3600 | 2772292 | 2836895.1 | 432882.1 | 553537.3 | 1850475.7 | 2731674.4 | 1.5% | 0.0% |
| sp(90;9;3;10) | >3600 | - | - | - | - | - | - | - | >3600 | 1303158.5 | 1371293.3 | 544018.3 | 494984.2 | 332290.7 | 1211453.9 | 7.6% | 0.0% |
| sp(100;10;2;15) | >3600 | - | - | - | - | - | - | - | >3600 | 3298717 | 3365603.9 | 406180.9 | 964688.6 | 1994734.4 | 3265726.8 | 1.0% | 0.0% |
| sp(110;11;3;12) | >3600 | - | - | - | - | - | - | - | >3600 | 1721897.8 | 1801231.7 | 658654.8 | 733895.2 | 408681.7 | 1368119.5 | 25.9% | 4.9% |
| sp(120;12;3;15) | >3600 | - | - | - | - | - | - | - | >3600 | 1948604.4 | 2046573.1 | 737950.9 | 865279.7 | 443342.4 | 1473377.1 | 32.3% | 6.1% |
| sp(150;15;4;20) | >3600 | - | - | - | - | - | - | - | >3600 | 2832831.4 | 2963761.3 | 680139 | 1434960.1 | 848662.2 | 1785391 | 58.7% | 22.2% |
| sp(160;15;4;35) | >3600 | - | - | - | - | - | - | - | >3600 | 1856191.7 | 2112701 | 747616.9 | 973153.4 | 391930.7 | 1509614.7 | 23.0% | 8.5% |
| sp(180;16;5;15) | >3600 | - | - | - | - | - | - | - | >3600 | 11225827.4 | 11369947.2 | 517282.7 | 516121.3 | 10336543.2 | 1583827.8 | 608.8% | 394.3% |
| sp(190;18;5;40) | >3600 | - | - | - | - | - | - | - | >3600 | 8433567.9 | 8577115.8 | 933526.3 | 1127668.6 | 6515920.9 | 1724431.2 | 389.1% | 315.2% |
| sp(200;15;4;20) | >3600 | - | - | - | - | - | - | - | >3600 | 7013669.2 | 7157953.5 | 722939.2 | 1618979.7 | 4816034.5 | 2112204.7 | 232.1% | 125.6% |
| sp(200;20;5;25) | >3600 | - | - | - | - | - | - | - | >3600 | 11596575.1 | 11648103.3 | 955143.2 | 1024996.1 | 9667964 | 1804904.1 | 542.5% | 417.0% |
| sp(230;18;4;50) | >3600 | - | - | - | - | - | - | - | >3600 | 16694178.3 | 16770022.4 | 643605.6 | 530508.4 | 15595908.5 | 2288106.2 | 629.6% | 572.9% |
| sp(250;20;5;40) | >3600 | - | - | - | - | - | - | - | >3600 | 19038852.3 | 19109855.6 | 568111.4 | 453226.7 | 18088517.4 | 2260729.7 | 742.2% | 538.5% |
| sp(280;25;5;60) | >3600 | - | - | - | - | - | - | - | >3600 | 17147524.3 | 17300515.5 | 821099.9 | 1339650.5 | 15139765.1 | 2718121.4 | 530.9% | 417.3% |
| sp(300;20;5;50) | >3600 | - | - | - | - | - | - | - | >3600 | 23590595.7 | 23713001.4 | 664023.3 | 687018.3 | 22361959.8 | 3135208.3 | 652.4% | 569.3% |
| sp(350;25;7;60) | >3600 | - | - | - | - | - | - | - | >3600 | - | - | - | - | - | - | - | - |
| sp(400;30;8;60) | >3600 | - | - | - | - | - | - | - | >3600 | - | - | - | - | - | - | - | - |

**Table 4.** Comparisons of the results obtained by the proposed fix and relax heuristics

| Sample problems | BBFR | | | | EBFR | | | | IF objective function |
|---|---|---|---|---|---|---|---|---|---|
| | Time (s) | Objective function | GAP% | RPD1% | Time (s) | Objective function | GAP% | RPD1% | |
| sp(18;5;2;3) | 5 | 1085419.9 | 0.0% | 7.8% | 4 | 1085419.9 | 0.0% | 7.8% | 1085419.9 |
| sp(30;5;2;5) | 7 | 1573106.5 | 0.0% | 0.2% | 8 | 1573106.5 | 0.0% | 0.2% | 1573106.5 |
| sp(50;8;2;5) | 11 | 1996220 | 1.9% | 6.0% | 13 | 1986683.9 | 1.4% | 5.5% | 1985395.7 |
| sp(80;8;2;10) | 28 | 2795922.3 | 2.4% | 0.9% | 27 | 2774833.6 | 1.6% | 0.1% | 2772292 |
| sp(90;9;3;10) | 92 | 1316478.3 | 8.7% | 1.0% | 85 | 1303702.0 | 7.6% | 0.0% | 1303158.5 |
| sp(100;10;2;15) | 146 | 3352710.9 | 2.7% | 1.6% | 168 | 3302512.7 | 1.1% | 0.1% | 3298717 |
| sp(110;11;3;12) | 217 | 1668596.7 | 22.0% | 1.6% | 305 | 1641834.8 | 20.0% | 0.0% | 1721897.8 |
| sp(120;12;3;15) | 297 | 1884103.2 | 27.9% | 2.6% | 322 | 1836895 | 24.7% | 0.0% | 1948604.4 |
| sp(150;15;4;20) | 601 | 2367881.9 | 32.6% | 2.2% | 742 | 2317363.4 | 29.8% | 0.0% | 2832831.4 |
| sp(160;15;4;35) | 588 | 1734803.6 | 14.9% | 1.4% | 940 | 1710683 | 13.3% | 0.0% | 1856191.7 |
| sp(180;16;5;15) | 733 | 2335106.1 | 47.4% | 2.8% | 836 | 2271283 | 43.4% | 0.0% | 11225827.4 |
| sp(190;18;5;40) | 852 | 2076319.4 | 20.4% | 2.2% | 1050 | 2031226.2 | 17.8% | 0.0% | 8433567.9 |
| sp(200;15;4;20) | 886 | 3197971.8 | 51.4% | 2.8% | 1217 | 3109355.2 | 47.2% | 0.0% | 7013669.2 |
| sp(200;20;5;25) | 1740 | 2312822.6 | 28.1% | 3.1% | 2380 | 2242845.8 | 24.3% | 0.0% | 11596575.1 |
| sp(230;18;4;50) | 2346 | 2509610.7 | 9.7% | 1.1% | 2741 | 2481078.3 | 8.4% | 0.0% | 16694178.3 |
| sp(250;20;5;40) | 2012 | 3096185.3 | 37.0% | 3.8% | 2616 | 2981688.5 | 31.9% | 0.0% | 19038852.3 |
| sp(280;25;5;60) | 2517 | 3376399.6 | 24.2% | 1.8% | 3381 | 3315070.8 | 22.0% | 0.0% | 17147524.3 |
| sp(300;20;5;50) | 2863 | 3544308.4 | 13.0% | 0.6% | 3422 | 3524570.8 | 12.4% | 0.0% | 23590595.7 |
| sp(350;25;7;60) | 3213 | 3978163.4 | - | 0.0% | >3600 | - | - | - | - |
| sp(400;30;8;60) | 3504 | 4907217.2 | - | 0.0% | >3600 | - | - | - | - |

**Table 5.** Comparisons of the results obtained by the proposed SAA and the optimal objective function

| Sample problems | SAA | | | | SBF objective function |
|---|---|---|---|---|---|
| | Time (s) | Objective function (best upper bound) | GAP% | RPD1% | |
| sp(18;5;2;3) | 12 | 1006942.1 | 0.0% | 0.0% | 1006942.1 |
| sp(30;5;2;5) | 48 | 1569669.2 | 0.0% | 0.0% | 1569669.2 |
| sp(50;8;2;5) | 117 | 1883116.6 | 0.1% | 0.0% | 1883116.6 |
| sp(80;8;2;10) | 460 | 2776958.6 | 137.1% | 0.2% | 3946716.4 |
| sp(90;9;3;10) | 866 | 1383745.7 | - | 6.2% | - |
| sp(100;10;2;15) | 1327 | 3603542.6 | - | 9.2% | - |
| sp(110;11;3;12) | 2755 | 1736546.6 | - | 5.8% | - |
| sp(120;12;3;15) | 3184 | 1914754.6 | - | 4.2% | - |
| sp(150;15;4;20) | >3600 | 2529716.2 | - | 9.2% | - |
| sp(160;15;4;35) | >3600 | 1857173 | - | 8.6% | - |
| sp(180;16;5;15) | >3600 | 2538216.6 | - | 11.8% | - |
| sp(190;18;5;40) | >3600 | 2353478.7 | - | 15.9% | - |
| sp(200;15;4;20) | >3600 | 3615740.7 | - | 16.3% | - |
| sp(200;20;5;25) | >3600 | 3706389 | - | 65.3% | - |
| sp(230;18;4;50) | >3600 | 4354958.7 | - | 75.5% | - |
| sp(250;20;5;40) | >3600 | - | - | - | - |
| sp(280;25;5;60) | >3600 | - | - | - | - |

The study so far leads to the following two questions: which method is doing the best in absolute terms? Which method outperforms in terms of different problem sizes? They are investigated in Figures 3 and 4, respectively. These figures show comparative graphs of the results obtained from Tables 4 and 5 associated with three proposed solution methods, BBFR, EBFR, and SAA, and the two criteria of the objective function value. As we can observe it in Figure 3, all methods achieve almost similar objective function values for small and medium-sized instances, whereas these values are larger for SAA method when the problem size has an increase. Likewise, Figure 4 compares the solution quality of the proposed methods using the RPD2 metric, where lower values indicate better performance. For small and medium-sized instances, all methods achieve RPD2 values close to zero, indicating similar performance. As problem size increases, the SAA method shows a clear deterioration, with RPD2 values rising sharply for large instances and reaching values above 30%. In contrast, both BBFR and EBFR maintain low, mostly negative RPD2 values across almost all instances, demonstrating stable and scalable performance. Among them, EBFR consistently provides the best results, particularly for larger instances. Overall, the figure confirms the superiority of the proposed fix-and-relax heuristics over the SAA approach in terms of robustness and solution quality for large-scale problems.

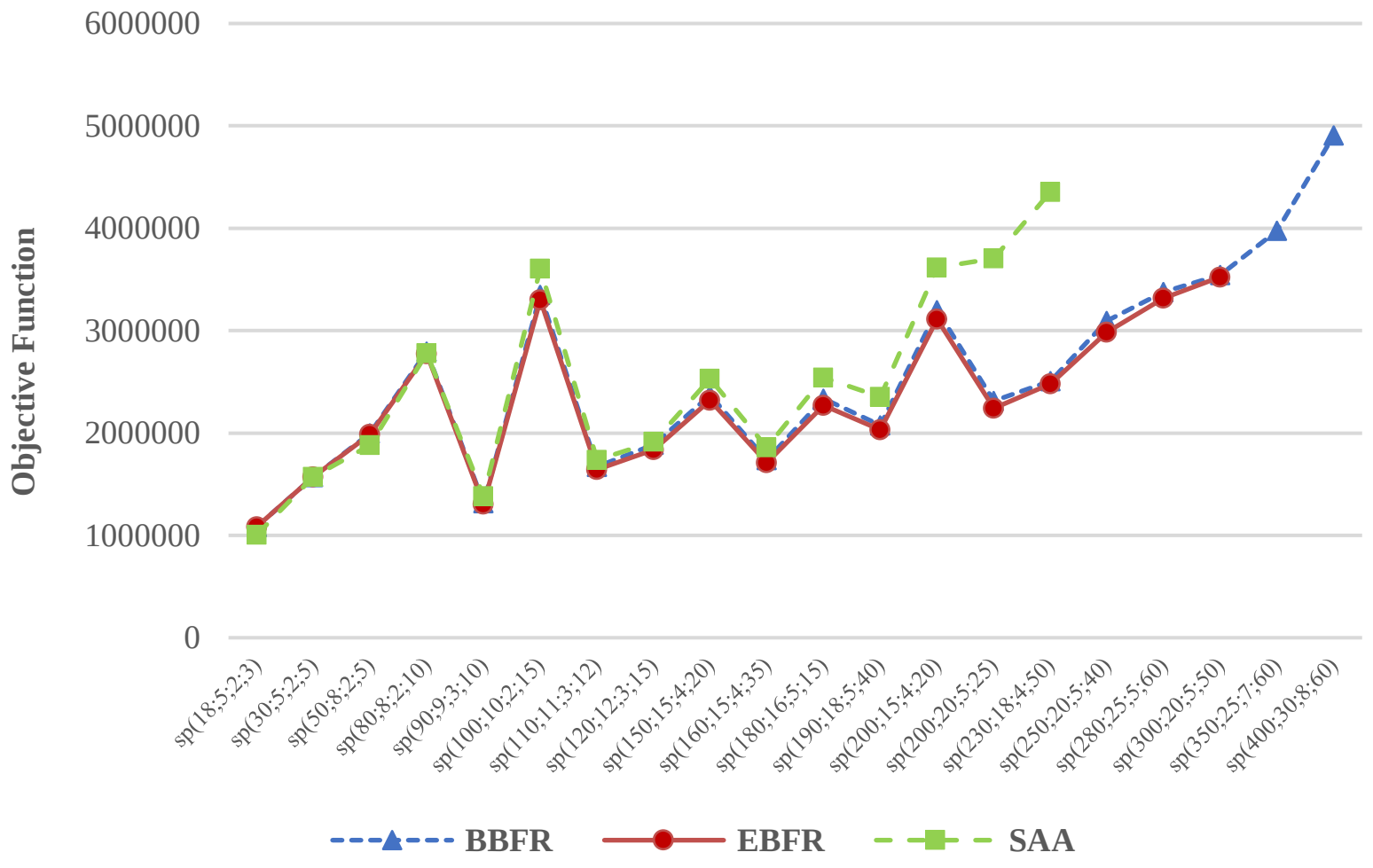

**Figure 3.** Comparison of the proposed solving methods regarding the objective function value

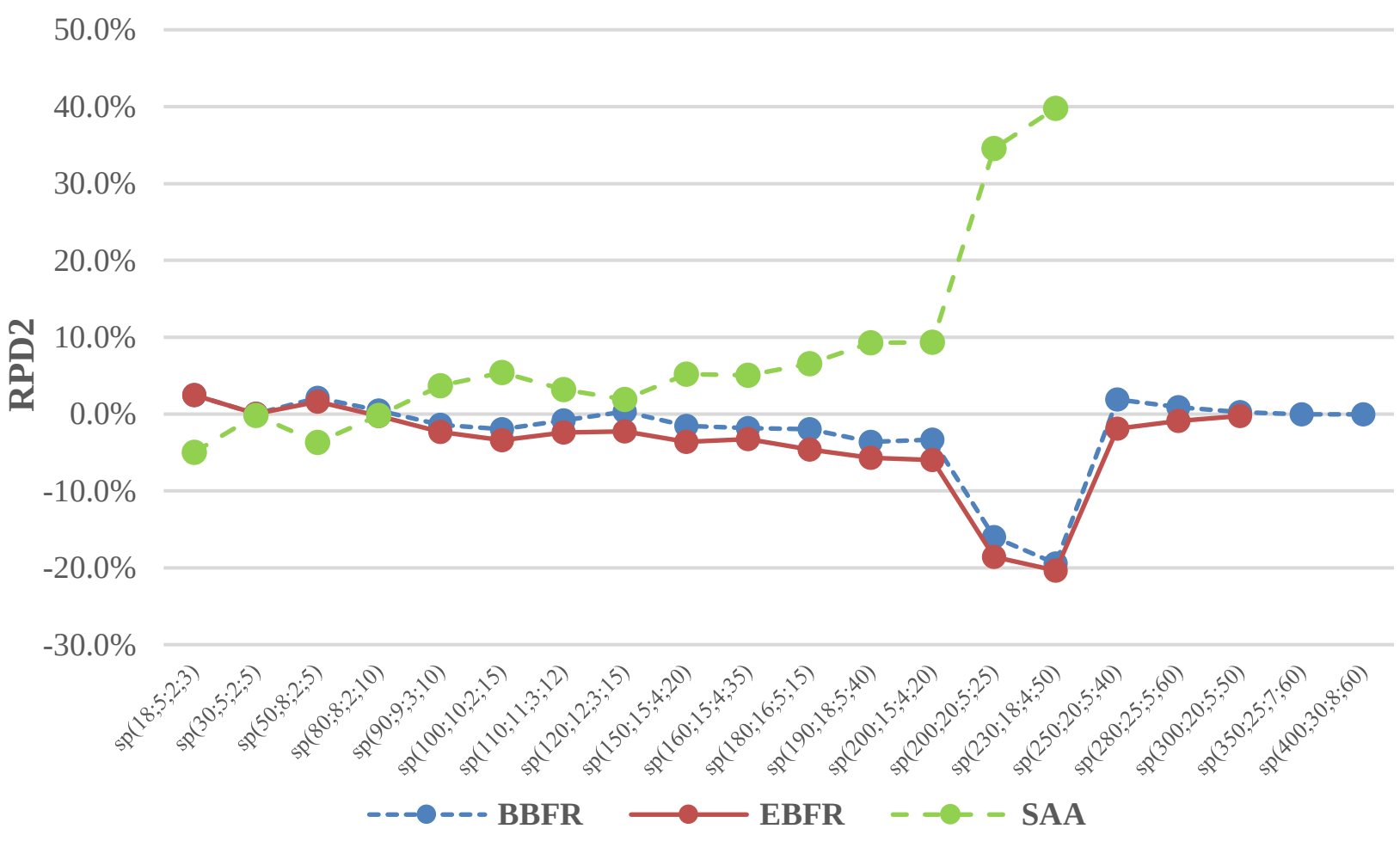

**Figure 4.** Comparison of the proposed solving methods regarding the RPD2 value

Having the analysis of BBFR, EBFR, and SAA in the previous tables and figures, let us now shift our concentration to the learned linear constraints for the situation in which mobile facility capacity should comply with disruption. Figure 5 assists us in the evaluation of five trained linear classifiers that can replace the chance Constraint (39) in the IF. In the context of the VTE-SCNDP, it somehow guarantees that the client assignment patterns to mobile facilities satisfy the 95% capacity-protection threshold. In this figure, each column represents a data set or a different size of the problem. Each row also represents a classifier to have five classifiers, including Logistic Regression (LR), Logistic Regression with L1 regularization (LR L1) where the less important features are zero, SGD Classifier, Perception, and Logistic Regression with a hyperparameter C equals 0.1 (LR C0.1) to control the strength of the regularization. Then, each one of 15 confusion matrices can show the accuracy of a classifier considering a particular case of the maximum number of regular facilities in the first echelon ($\gamma_1$) and the capacity of a mobile facility ($MC_k$). We assume that a high true-positive rate (at least 90%) and low false-positive rates for any matrix indicates that the classifiers reliably distinguish feasible assignment patterns (satisfy the 95% capacity requirement) from infeasible ones that violate it. This analysis reveals problem settings in which a classifier may misclassify assignment patterns, enabling the selection of reliable classifiers and reducing the risk of solutions that violate the predefined service level.

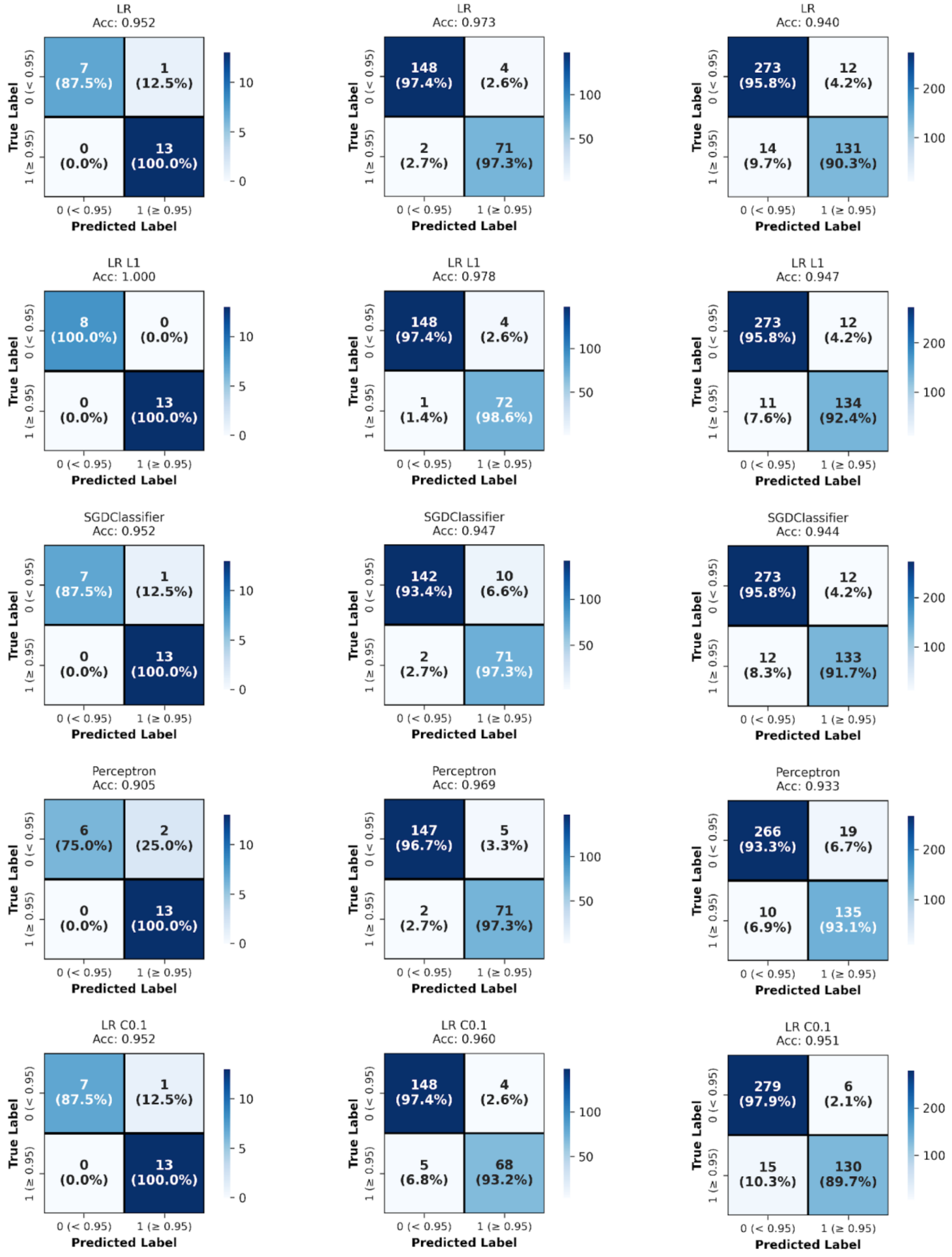

a) $\gamma_1 = 3;\ MC_k = 3$ b) $\gamma_1 = 10;\ MC_k = 4$ c) $\gamma_1 = 10;\ MC_k = 5$

**Figure 5.** Confusion matrices illustrating linear classifiers performance with different $\gamma_1$ and $MC_k$

**Table 6.** Performance of machine learning classifiers

| Model | Parameters | Accuracy | Precision | Recall | F1-score |
|---|---|---|---|---|---|
| Logistic Regression | L2, C=1.0 | 95.24% | 92.86% | 100.00% | 96.30% |
| L1 Logistic Regression | L1, C=1.0 | 100.00% | 100.00% | 100.00% | 100.00% |
| SGD (Linear SVM) | hinge, max_iter=1000 | 95.24% | 92.86% | 100.00% | 96.30% |
| Perceptron | sklearn default, max_iter=1000 | 90.48% | 86.67% | 100.00% | 92.86% |
| Logistic Regression (C=0.1) | L2, C=0.1 | 95.24% | 92.86% | 100.00% | 96.30% |

Consider Figure 5 further, we should verify the credibility and robustness of the learned constraints in order to be used instead of the intractable chance constraint of the model. For example, let's focus on the combination of $\gamma_1 = 10$ and $MC_k = 5$ in the last column of the figure. Then, it is easy to observe that LR C0.1 has the highest accuracy of 0.951 in comparison with other classifiers with accuracies of 0.940, 0.947,0.944, and 0.933. Perhaps, the reason behind this high accuracy of LR C0.1 classifier for the case of $\gamma_1 = 10$ and $MC_k = 5$ is that the parameter C equal to 0.1 can decrease overfitting in order to improve the accuracy of the logistic regression for large instances. This can justify the integration of such a learned intercept and the coefficients of the classifier into the IF.

Table 6 reports the performance of the evaluated classification models for a representative dataset with $\gamma_1 = 5$ and $MC_k = 3$, and regular facility failure probability equal to 0.15 in the first echelon. The metrics include accuracy (overall correctness), precision (proportion of predicted feasible patterns that are truly feasible), recall (proportion of truly feasible patterns correctly identified), and F1-score (a combined measure balancing precision and recall). All models achieve high performance, indicating that the feasibility of assignment patterns can be effectively learned. In particular, recall is equal to 100% for all models, implying that no false negatives occur, i.e., no truly feasible pattern is incorrectly classified as infeasible. However, models with precision below 100% exhibit a small number of false positives, meaning that some infeasible patterns may be incorrectly classified as feasible, potentially leading to slight violations of the 95% service level. Among the tested models, L1-regularized logistic regression achieves perfect performance across all metrics and is therefore selected for this dataset. It should be noted, however, that the best-performing model may vary across different datasets, and the model selection procedure is applied independently for each instance.

**Table 7.** Improvement in CPU time of SAA with merging similar scenarios

| Sample problems | SAA with merging similar scenarios | SAA without merging similar scenarios | Time improvement (%) |
|---|---|---|---|
| | Time (s) | Time (s) | |
| sp(18;5;2;3) | 12 | 25 | 52% |
| sp(30;5;2;5) | 48 | 682 | 93% |
| sp(50;8;2;5) | 117 | >3600 | - |
| sp(80;8;2;10) | 460 | >3600 | - |
| sp(90;9;3;10) | 866 | >3600 | - |
| sp(100;10;2;15) | 1327 | >3600 | - |
| sp(110;11;3;12) | 2755 | >3600 | - |
| sp(120;12;3;15) | 3184 | >3600 | - |
| sp(150;15;4;20) | >3600 | >3600 | - |

Table 7 shows the evaluation of the effect of reducing the number of scenarios by merging similar scenarios in SAA. This strategy considers a set of scenarios as redundant ones after determining the value of variables $y_{l,j}$. It is due to the fact that they involve identical states for the subset of opened facilities. Therefore, we can decrease the number of second-stage subproblems that require to be solved by a simple consolidation of identical states. The results indicate that applying this approach within the SAA significantly reduces CPU time, with the average decreasing from 2878 s to 1374 s. More precisely, Table 7 shows that merging steps decreases CPU time to 52% and 93% for the first two instances. For medium-sized problems such as sp(50;8;2;5) and sp(80;8;2;10), SAA with merging can converge in a reasonable time instead of failing to find any feasible solutions. Hence, one can argue that technique is vital for converting SAA into a computationally efficient approach for large-scale VTE-SCNDPs.

Let us now confine ourselves to the sensitivity analysis in order to evaluate how a scenario performs if a major factor fluctuates in practice. Tables 8 and 9 depict each method's sensitivity to disruption to present a practice-focused view of the methods. Regarding Table 8, it shows the performance results of BBFR and EBFR under the influence of changes in the failure probabilities of the facilities in the sample sp(80;8;2;10). The table shows that EBFR has a sustainable performance under different scenarios of failure probability and provides an optimal solution in all cases. In the BBFR method, except for $p_1 = p_2 = 0.9$, where we see a significant drop in quality (RPD1 = 19.32%), we do not notice any significant changes in the performance of this method in other failure probabilities. The average RPD1, regardless of $p_1 = p_2 = 0.9$, is 1.35%.

**Table 8.** Performance of the BBFR and EBFR as a result of changes in the failure probability

| Failure probability | BBFR | | EBFR | | IF |
|---|---|---|---|---|---|
| | Objective function | RPD1 | Objective function | RPD1 | Objective function |
| 0.1 | 2725733.4 | 0.25% | 2718936.1 | 0.00% | 2718936.1 |
| 0.2 | 2796099.6 | 0.57% | 2780252.2 | 0.00% | 2780252.2 |
| 0.3 | 2851373.1 | 1.82% | 2800405.8 | 0.00% | 2800405.8 |
| 0.4 | 2926463.6 | 2.65% | 2850914.4 | 0.00% | 2850914.4 |
| 0.5 | 2949847.1 | 1.32% | 2911416.5 | 0.00% | 2911416.5 |
| 0.6 | 2943357.5 | 0.00% | 2943357.5 | 0.00% | 2943357.5 |
| 0.7 | 3044387.2 | 0.00% | 3044387.2 | 0.00% | 3044387.2 |
| 0.8 | 3510732.2 | 4.65% | 3354737.0 | 0.00% | 3354737.0 |
| 0.9 | 4324918.8 | 19.32% | 3624638.7 | 0.00% | 3624638.7 |
| 1.0 | 3826129.4 | 0.85% | 3793881.5 | 0.00% | 3793881.5 |

Table 9 shows how the SAA method functions as a result of changes in the failure probabilities of sp(50:8:2:5). The table shows that increasing the probabilities of full failure reduces the performance of this method. Hence, the RPD1 value for $p_1 = p_2 = 0.9$ is 27.65%. Also, the slope of the drop in the quality of the solutions obtained by this method increases with increased probabilities of full failure of the facilities. In the SAA method, the average RPD1 value for variations in the range 0.1 to 0.7 is equal to 0.46%, whereas the average of this value for variations in the interval 0.8 to 0.9 is equal to 18.71%.

**Table 9.** Performance of the SAA as a result of changes in the failure probability

| Failure probability | SAA | | SBF |
|---|---|---|---|
| | Objective function | RPD1 | Objective function |
| 0.1 | 1866620.5 | 0.00% | 1866620.5 |
| 0.2 | 2420934.1 | 0.00% | 2420934.1 |
| 0.3 | 2864960.5 | 0.00% | 2864960.5 |
| 0.4 | 2565273.4 | 1.45% | 2528565.3 |
| 0.5 | 3080710.8 | 1.75% | 3027651.6 |
| 0.6 | 3156520.7 | 0.00% | 3156520.7 |
| 0.7 | 3640423.2 | 0.00% | 3640423.2 |
| 0.8 | 3887399.4 | 9.76% | 3541631.3 |
| 0.9 | 3721922.8 | 27.65% | 2913303.4 |
| 1.0 | 3354397.6 | 0.00% | 3354397.6 |

Beyond algorithmic performance in the tables above, our results yield strategic insights obtained from Table 10. More precisely, our models implement an integrated optimization across all echelons to shape it as a cross-echelon coordination. Establishing this assumption requires cooperation between the owners of all echelons of the network. Table 10 examines the effectiveness of this cooperation. This table shows the results obtained from solving the model in the two approaches: the integrated optimization and the hierarchical optimization. Each echelon is independently optimized from other echelons in the hierarchical approach. At first, the lowest echelon of the network (echelon 1) is optimized, and its results are fixed. Next, the second echelon is optimized, and its results are fixed. According to the results of Table 10, the priority of the integrated approach over the hierarchical approach is evident. The average RPD1 criterion in this approach is 11.78% better than the hierarchical approach.

**Table 10.** Comparison of the results obtained from two integrated and hierarchical optimization

| Sample problems | Integrated optimization | | hierarchical optimization | |
|---|---|---|---|---|
| | RPD1% | Obj | RPD1% | Obj |
| sp(18;5;2;3) | 0.0% | 1085419.9 | 15.22% | 1250620.8 |
| sp(30;5;2;5) | 0.0% | 1573106.5 | 7.41% | 1689673.7 |
| sp(50;8;2;5) | 0.0% | 1986683.9 | 10.32% | 2191709.7 |
| sp(80;8;2;10) | 0.0% | 2774833.6 | 6.87% | 2965464.7 |
| sp(90;9;3;10) | 0.0% | 1303702.0 | 13.32% | 1477355.1 |
| sp(100;10;2;15) | 0.0% | 3302512.7 | 10.85% | 3660835.3 |
| sp(110;11;3;12) | 0.0% | 1641834.8 | 18.47% | 1945081.7 |

In short, the experiments in this section could help us gain some insights for network design with facility disruption. Our observation is that practitioners can prioritize the IF with the EBFR and the SAA method by merging similar scenarios. Then, considering failure probabilities in the first columns of Tables 8 and 9, these robust methods put confidence in practice. More importantly, we can reach viability under disruption if we implement an integrated version of the network design. Then, one can argue that aforementioned outcomes result in the long-term survivability of two-echelon supply chains. The reason behind this intuition is that it minimizes the costs across all disruption scenarios (e.g., the objective function in Equation (1)) due to proactive coordination of backup options in two echelons. In other words, the integration allows for products and information flow between the facilities in order to reach both local and global targets.

## 8. Conclusions

The focus of this study is on the design of a viable two-echelon supply chain network design problem (VTE-SCNDP) under facility disruptions. We have assumed that the network must maintain itself and survive in a changing environment, especially due to cross-echelon disruptions risks. We have developed two new formulations based on implicit and scenario-based formulations in order to add metrics such as resiliency, agility, and environmental impact to the design of such a network. The computational results show that the IF is a good alternative to the SBF in large-scale problems and provides better quality of solutions and running time. Based on computational results, the IF models and two custom-designed fix-and-relax heuristics are efficient for large-sized problems, which are intractable for the exact SBF. In addition, we have introduced a Sample Average Approximation algorithm for the SBF to improve its performance computationally for medium-sized problems.

We also employed machine learning techniques to meet challenges from probabilistic service-level constraints common in practice. To do so, learned linear classifiers are developed and trained on data to be used instead of the intractable chance constraint associated with mobile-facility capacity. After such modifications, the model could meet any predefined service level computationally practical. Specifically, we could narrow the theory-practice gap in order to make optimization and machine learning techniques applicable for VTE-SCNDPs.

Computational results have shown that the adoption of an integrated approach to decision making and simultaneous optimization at all echelons are much more effective than hierarchical optimization. Therefore, interaction and collaboration between owners of different echelons of the network is strongly recommended. Finally, the study also provided some insights into the VTE-SCNDPs to conclude that an integrated approach can provide an opportunity for decision-makers to decrease costs in comparison with hierarchical planning (e.g., due to its 11.7% superiority on average over the RPD1 criterion of our test instances).

Future studies could focus on the interdependency of reliability levels across echelons. In multi-echelon systems, different types of relationships between echelons could be explored, such as cases where the probability of failure or the cost of demand fulfillment depends on the planned reliability levels in other network tiers. While this study focuses on independent facility disruptions and their operational interdependencies, a vital extension would be to incorporate correlated disruption scenarios. Such scenarios, stemming from pandemics, regional conflicts, or economic crises, could model the causal cascading (ripple) effect, where a single event triggers simultaneous failure across multiple network nodes. Exploring how these complex dynamics influence the design of viable and resilient supply chains remains a promising direction for further investigation. Moreover, a natural extension of this work is to adapt the ML framework to handle continuous demand variables and rigid capacity limits, further demonstrating the versatility of the proposed approach in diverse industrial contexts. Finally, the use of other machine learning methods, such as ensemble ones for constraints approximation, could be a future research direction.

## Appendix A. Sample average approximation algorithm

In this section, we propose a sample average approximation (SAA) heuristic to solve the VTE-SCNDP. The SAA method is one of sampling-based procedures that has had a successful performance in solving stochastic problems with a very large number of scenarios, such as (Aydin and Murat, 2013; Schütz et al., 2009; Verweij et al., 2003). In the SAA approach, a number of sample scenarios are selected randomly, and the true expected value function of the stochastic problem is estimated by an approximate problem that is constructed using these sampled scenarios. In our case, the SAA formulation is stated as (A.1):

$$\begin{aligned} min \sum_{l}\sum_{j\le J_l} s_{l,j} y_{l,j} + \sum_{s\in e_m}\frac{1}{N}\sum_{k\le K}\delta_k\,\mu_{k,s} \\ + \sum_{s\in e_m}\frac{1}{N}\left(\sum_{i\le I_1} d_i\left(\sum_{j\le J_1} c_{1,i,j}\,X_{1,i,j,s} + \sum_{k\le K} c'_{i,k}\,M_{i,k,s}\right) + \sum_{i\le I_2}\sum_{j\le J_2} c_{2,i,j}\,u_{i,j,s}\right) + \\ \sum_{s\in e_m}\frac{1}{N}\left(\sum_{i\le I_1} h_{1,i} d_i\,Z_{1,i,s} + \sum_{i\le I_2} h_{2,i} v_{i,s}\right) \end{aligned} \tag{A.1}$$

$s.t.$

Constraint (2) to (13) $\quad s \in e_m$

$$\sum_{s\in e_m}\frac{1}{N}\left(\sum_{i\le I_1} d_i\,e_1\left(\sum_{j\le J_1} f_{1,i,j}\,X_{1,i,j,s} + \sum_{k\le K} f'_{kj}\,M_{i,k,s}\right) + \sum_{i\le I_2}\sum_{j\le J_2} f_{2,i,j}\,u_{i,j,s} e_2\right) \le E_{max} \tag{A.2}$$

The above model is solved $M$ times ($m = 1, \dots, M$) for $M$ samples $\{e_1, e_2, \dots, e_M\}$ consisting of $N$ random scenarios $e_m = \{s_m^1, s_m^2, \dots, s_m^N\}$. For each $m = 1, \dots, M$, let $y^m$and $Z^m$ denote the first-stage vector solution ($y_{l,j}$) and the optimal objective function value of the SAA formulation (Objective function (A.1) and Constraints (2) to (13) and (A.2)). After solving the SAA problem $M$ times and obtaining $M$ first-stage solutions $y^1, \dots, y^M$ with associated objective values $Z^1, \dots, Z^M$, the vector with the smallest objective (shown in Equation (A.3)) is selected for full second-stage evaluation,

$$y^{\text{best}} = \arg \min_{m=1,\dots,M} \mathbb{Z}^m \tag{A.3}$$

In principle, the second-stage evaluation of a given $y^m$ requires solving the original two-stage model (Objective (1) with Constraints (2) – (14)) for all original scenarios. However, direct evaluation is computationally challenging because Constraint (14) couples all scenarios through a single expected-emission limit, preventing decomposition into smaller subproblems.

To overcome this difficulty, we apply a Lagrangian relaxation of Constraint (14) and group scenarios into small sets. Each group of scenarios is then solved as a separate subproblem, while the dual variable ensures the global emission limit is respected across all groups. This decomposition drastically reduces computational complexity while maintaining feasibility with respect to the constraint on the environmental impact. Let $\hat{Z}^{best}$ denote the second-stage objective value obtained for $y^{best}$ under this LR-based decomposition. This value provides an upper bound on the true optimum and is the output of the SAA heuristic.

The possibility of reducing the number of scenarios by merging similar scenarios is another advantage that can be obtained by the two-stage solution of the model. Determining the values of the variables $y_{l,j}$ in the first stage makes it possible to reduce the number of problem scenarios by merging similar scenarios before solving the MIP model that remains in the second stage. With this reduction, the number of variables and constraints of the model are directly reduced, which results in reduced computational complexity. Table A.1 shows the reduction of the number of scenarios from 16 to 4 for a problem with 4 facilities ($j_1$ to $j_4$), where each facility has two random states, one active and one inactive. This table shows that if facilities 1 and 2 are established and facilities 3 and 4 remain closed in the first-stage decisions, the total number of scenarios can be reduced from 16 to 4 by combining scenarios 1 to 4; 5 to 8; 9 to 12; and 13 to 16.

**Table A.1.** Reducing the number of scenarios from 16 to 4 by merging similar scenarios

| $j_1$ | $j_2$ | $j_3$ | $j_4$ | $p(a)$ | $j_1$ | $j_2$ | $p(a')$ |
|---|---|---|---|---|---|---|---|
| Inact. | Inact. | Inact. | Inact. | $p(a_1)$ | Inact. | Inact. | $p(a_1) + p(a_2) + p(a_3) + p(a_4)$ |
| Inact. | Inact. | Inact. | Act. | $p(a_2)$ | | | |
| Inact. | Inact. | Act. | Inact. | $p(a_3)$ | | | |
| Inact. | Inact. | Act. | Act. | $p(a_4)$ | | | |
| Inact. | Act. | Inact. | Inact. | $p(a_5)$ | Inact. | Act. | $p(a_5) + p(a_6) + p(a_7) + p(a_8)$ |
| Inact. | Act. | Inact. | Act. | $p(a_6)$ | | | |
| Inact. | Act. | Act. | Inact. | $p(a_7)$ | | | |
| Inact. | Act. | Act. | Act. | $p(a_8)$ | | | |
| Act. | Inact. | Inact. | Inact. | $p(a_9)$ | Act. | Inact. | $p(a_9) + p(a_{10}) + p(a_{11}) + p(a_{12})$ |
| Act. | Inact. | Inact. | Act. | $p(a_{10})$ | | | |
| Act. | Inact. | Act. | Inact. | $p(a_{11})$ | | | |
| Act. | Inact. | Act. | Act. | $p(a_{12})$ | | | |
| Act. | Act. | Inact. | Inact. | $p(a_{13})$ | Act. | Act. | $p(a_{13}) + p(a_{14}) + p(a_{15}) + p(a_{16})$ |
| Act. | Act. | Inact. | Act. | $p(a_{14})$ | | | |
| Act. | Act. | Act. | Inact. | $p(a_{15})$ | | | |
| Act. | Act. | Act. | Act. | $p(a_{16})$ | | | |

For any feasible solution $(y^m | m = 1, \dots, M)$ obtained from solving the SAA formulation, statistical upper bound estimation can be obtained by sampling $N'$ ($N' \gg N$) scenarios $E = \left\{s^1, s^2, \dots, s^{N'}\right\}$ and solving these second-stage problem, while the objective function of second-stage problem is as (A.4):

$$min \sum_{l} \sum_{j \le J_l} s_{l,j} y_{l,j}^m + \sum_{s \in E} \frac{1}{N'} \sum_{k \le K} \delta_k \, \mu_{k,s} + \sum_{s \in E} \frac{1}{N'} \left( \sum_{i \le I_1} d_i \left( \sum_{j \le J_1} c_{1,i,j} \, X_{1,i,j,s} + \sum_{k \le K} c'_{i,k} \, M_{i,k,s} \right) + \sum_{i \le I_2} \sum_{j \le J_2} c_{2,i,j} \, u_{i,j,s} \right) + \sum_{s \in e_m} \frac{1}{N'} \left( \sum_{i \le I_1} h_{1,i} d_i \, Z_{1,i,s} + \sum_{i \le I_2} h_{2,i} v_{i,s} \right) \tag{A.4}$$

Also, a lower bound for the true optimum is provided by the expected value of SAA function (Mak et al., 1999). Therefore, the average of the $Z^m$ values, i.e., $\overline{\mathbb{Z}}^M = \left( {\sum_m \mathbb{Z}^m}/{M} \right)$ provides a statistical estimate of a lower bound for the optimal value. We can obtain an estimate for the gap by calculating the difference between the estimated upper bound and lower bound.

**Acknowledgement:**

This research has been supported by the European Union under the project ROBOPROX (reg. no. CZ.02.01.01/00/22008/0004590) and the Grant Agency of the Czech Republic under the Project GACR 25-17904S.